\documentclass[a4paper,10pt,english]{article}
\usepackage{amsmath,amssymb,graphicx} 
\usepackage{authblk}
\usepackage{epsfig}
\usepackage{pstricks}
\usepackage{multirow}
\usepackage[bookmarks]{hyperref}
\usepackage{graphicx}
\usepackage{fancybox}
\usepackage{amssymb}
\usepackage{amsbsy}
\usepackage{amsmath}
\usepackage{fancyhdr}
\usepackage{mathrsfs}
\usepackage{subfigure}
\usepackage{pifont}
\usepackage{colortbl}
\usepackage{pgf}
\usepackage{pstricks}
\usepackage{epsfig}
\usepackage{booktabs}
\usepackage{multirow}
\usepackage{axodraw4j}
\usepackage{fmtcount}
\usepackage{booktabs}
\usepackage{graphicx}
\usepackage{hyperref}
\usepackage{colortbl}
\usepackage{longtable}
\usepackage{cite}

\usepackage[ a4paper, left=2cm, right=2cm, top=2.5cm, bottom=2.5cm]{geometry}

\def\MET{{E\!\!\!\!\slash_{T}}}

\def\pslash{\not{\hbox{\kern-4pt $p$}}}
\def\qslash{\not{\hbox{\kern-4pt $q$}}}
\def\lv{\not{\hbox{\kern-4pt $L$}}}

\begin{document}


\begin{flushright}
{KEK-TH-1560}\\
{LYCEN 2012-04 }\\
\end{flushright}
\vskip0.1cm\noindent
\begin{center}
{{\large {\bf LHC signatures of vector-like quarks }}
\\[0.5cm]
{\large Yasuhiro Okada$^{1,2}$ and Luca Panizzi$^3$}\\[0.30 cm]
{\it $^1$KEK Theory Center, Institute of Particle and Nuclear Studies, KEK} \\
{\it 1-1 Oho, Tsukuba, Ibaraki 305-0801, Japan.}\\[0.20 cm]
{\it $^2$Department of Particle and Nuclear Physics, Graduate University for Advanced Studies (Sokendai),}\\
{\it 1-1 Oho, Tsukuba, Ibaraki 305-0801, Japan.}\\[0.20 cm]
{\it $^3$Universit\'e de Lyon, France; Universit\'e Lyon 1, CNRS/IN2P3,}\\
{\it UMR5822 IPNL, F-69622 Villeurbanne Cedex, France.}
}
\\[1.25cm]
\end{center}

\begin{abstract}
This work provides an overview on the current status of phenomenology and searches for heavy vector-like quarks, which are predicted in many models of new physics beyond the Standard Model. Searches at Tevatron and at the LHC, here listed and shortly described, have not found any evidence for new heavy fermionic states (either chiral or vector-like), and have therefore posed strong bounds on their masses: depending on specific assumptions on the interactions and on the observed final state, vector-like quarks with masses up to roughly 400-600 GeV have been excluded by all experiments. In order to be as simple and model-independent as possible, the chosen framework for the phenomenological analysis is an effective model with the addition of a vector-like quark representation (singlet, doublet or triplet under $SU(2)_L$) which couples through Yukawa interactions with all SM families. The relevance of different observables for the determination of bounds on mixing parameters is then discussed and a complete 
overview of possible two body final states for every vector-like quark is provided, including their subsequent decay into SM particles. A list and short description of phenomenological analyses present in literature is also provided for reference purposes. 
\end{abstract}

\tableofcontents

\section{Introduction}

A fermion is defined to be vector-like if its left- and right-handed chiralities belong to the same representation of the symmetry group $G$ of the underlying theory, where $G\equiv SU(3)_c \otimes SU(2)_L \otimes U(1)_Y$ in the Standard Model (SM). At present there is no evidence of the existence of vector-like quarks (VLQs), nevertheless they are key ingredients for many models of physics beyond the Standard Model (BSM) and there is a vast literature about their properties and phenomenology. Recent observations at the LHC strongly point towards the existence of a resonance which is compatible with a SM Higgs \cite{HiggsATLAS,HiggsCMS}, thus seeming to complete the SM picture. The putative Higgs discovery would leave unsolved many long-standing issues related to the nature of fermion mass hierarchies (Yukawas are completely free parameters) and to the Higgs mass stability itself (divergence of radiative corrections), which can probably be explained extending the SM to encompass new, yet undiscovered, states.
 Dedicated experimental searches for VLQs have already been undertaken and will be improved in the next future, therefore a detailed understanding of the properties of VLQs is essential to drive and tune forthcoming analyses. 

From a theoretical point of view, VLQs have been introduced in many different models; the most studied scenarios which predict the presence of VLQs can be divided into broad categories\footnote{A description of the various models as well as their consistency against the observations of the 125 GeV Higgs-like resonance is beyond the scopes of the present analysis; details can be found in the original works and references therein. Here, it is sufficient to note how the emergence of VLQs is a recurrent consequence in many models of BSM physics.}:
\begin{itemize}
\item{\it Composite Higgs Models:} the EW symmetry breaking is driven by a condensate of the top quark and a VL singlet involving a see-saw mechanism between the two states \cite{Dobrescu:1997nm,Chivukula:1998wd,Collins:1999rz,He:2001fz,Hill:2002ap,Contino:2006qr,Anastasiou:2009rv,Kong:2011aa,Carmona:2012jk,Gillioz:2012se};
\item{\it Extra Dimensions:} excited partners of SM quarks belonging to heavier tiers of universal extra-dimensional scenarios are vector-like;
\item{\it Gauging of the flavour group:} VL fermions are required for anomaly cancellation and can play a role in the mechanisms of quark mass generation \cite{Davidson:1987tr,Babu:1989rb,Grinstein:2010ve,Guadagnoli:2011id};
\item{\it Little Higgs Models:} VL states appear as partners of SM fermions in larger representations of the symmetry group \cite{ArkaniHamed:2002qy,Han:2003wu,Perelstein:2003wd,Schmaltz:2005ky,Carena:2006jx,Matsumoto:2008fq}.
\item{\it Supersymmetric non-minimal extensions of the SM:} VL matter can be introduced in non minimal supersymmetric models to increase corrections the Higgs mass without affecting too much EW precision observables \cite{Moroi:1991mg,Moroi:1992zk,Babu:2008ge,Martin:2009bg,Graham:2009gy,Martin:2010dc}, and it appears also in non-minimal, GUT-inspired, supersymmetric scenarios \cite{Kang:2007ib}.
\end{itemize}
VLQs can also appear in models which try to explain measured asymmetries in different processes.
\begin{itemize}
\item in \cite{Choudhury:2001hs,Kumar:2010vx} VLQs are introduced to explain the observed $A_{FB}^b$ asymmetry: bottom partners can mix with the bottom quark and induce modifications of its coupling with the Z boson. 
\item the forward-backward asymmetry $A_{FB}$ in top pair production, measured at Tevatron, can be explained with the existence of a color octet with a large decay width; this condition can be obtained if the color octet is allowed to decay to a heavy VL state and a SM fermion \cite{Barcelo:2011vk,Barcelo:2011wu}. 
\end{itemize}

From the phenomenological point of view, signatures of VLQs have been largely analysed in literature, both from a model independent perspective and within specific scenarios. The presence of flavour changing neutral currents, a distinctive feature of VLQs, leads to a wide range of possible final states, which have been (and will certainly be) analysed in detail in order to drive the experimental search of these new states.

The present study is organized as follows: in Sect.~\ref{Sect:model} the minimal extension of the SM with the presence of VLQs is described, together with an overview of observables which can provide bounds to the mixing parameters of the theory; Sect.~\ref{Sect:searches} contains a summary of all searches for heavy fermions at Tevatron and LHC, focusing on the assumptions that lie beneath the obtained bounds; in Sect.~\ref{Sect:signatures} a list of all allowed final states coming from VLQ resonant production in two-body intermediate states at the LHC is presented, considering all the possible decay channels for the different VLQ representations; moreover, specific signatures, analysed in literature, are shortly described; in Sect.~\ref{Sect:VLQcontributions} some examples of the possible contribution of VLQs to other observables is discussed.

Some of the results reported here were published in Refs.~\cite{Cacciapaglia:2010vn,Cacciapaglia:2011fx}.

\section{Model framework}
\label{Sect:model}

\subsection{Representations and couplings}

The minimal scenarios with the presence of VLQs besides SM particles are those in which the new states interact with SM quarks and the Higgs boson through Yukawa couplings. Classifying VLQs in multiplets of $SU(2)_L$, it is possible to write gauge-invariant interaction terms only for singlets, doublets and triplet representations. All the possibilities are shown in Tab.~\ref{VLrepresentations}.
\begin{table}[htb]
\begin{eqnarray*}
\setlength{\arraycolsep}{2pt}
\begin{array}{ccccccccccc}
&\multicolumn{3}{c}{\mbox{SM quarks}}&\multicolumn{2}{c}{\mbox{Singlets}}&\multicolumn{3}{c}{\mbox{Doublets}}&\multicolumn{2}{c}{\mbox{Triplets}}\\
& 
\begin{array}{c} ~ \\ \left(\begin{array}{c} u \\ d \end{array}\right) \\ ~ \end{array} &
\begin{array}{c} ~ \\ \left(\begin{array}{c} c \\ s \end{array}\right) \\ ~ \end{array} &
\begin{array}{c} ~ \\ \left(\begin{array}{c} t \\ b \end{array}\right) \\ ~ \end{array} &
\begin{array}{c} ~ \\ (U) \\ ~ \\ ~ \end{array} &
\begin{array}{c} ~ \\ ~ \\ (D) \\ ~ \end{array} & 
\begin{array}{c} \left(\begin{array}{c} X \\ U \end{array}\right) \\ ~ \\ ~ \end{array} &
\begin{array}{c} ~ \\ \left(\begin{array}{c} U \\ D \end{array}\right) \\ ~ \end{array} &
\begin{array}{c} ~ \\ ~ \\ \left(\begin{array}{c} D \\ Y \end{array}\right) \end{array} & 
\begin{array}{c} \left(\begin{array}{c} X \\ U \\ D \end{array}\right) \\ ~ \end{array} &
\begin{array}{c} ~ \\ \left(\begin{array}{c} U \\ D \\ Y \end{array}\right) \end{array} \\
\midrule
SU(2)_L & 
\multicolumn{3}{c}{\begin{array}{c}q_L=2\\q_R=1\end{array}} & 
\multicolumn{2}{c}{1} & 
\multicolumn{3}{c}{2} & 
\multicolumn{2}{c}{3} \\
\midrule
U(1)_Y  & 
\multicolumn{3}{c}{\begin{array}{c}q_L=1/6\\u_R=2/3\\d_R=-1/3\end{array}} & 2/3 & -1/3 & 7/6 & 1/6 & -5/6 & 2/3 & -1/3 \\
\midrule
\mathcal{L}_Y &
\multicolumn{3}{c}{\begin{array}{c}-y_u^i \bar q_L^i H^c u_R^i \\- y_d^i \bar q_L^i V_{CKM}^{i,j} H d_R^j\end{array}} &
\multicolumn{2}{c}{\begin{array}{c}-\lambda_u^i \bar q_L^i H^c U_R \\-\lambda_d^i \bar q_L^i H D_R \end{array}} &
\multicolumn{3}{c}{\begin{array}{c}-\lambda_u^i \psi_L H^{(c)} u_R^i \\-\lambda_d^i \psi_L H^{(c)} d_R^i \end{array}} &
\multicolumn{2}{c}{\begin{array}{c}-\lambda_i \bar q_L^i \tau^a H^{(c)} \psi_R^a \end{array}} \\
\midrule
\mathcal{L}_m & \multicolumn{3}{c}{\mbox{not allowed}} & \multicolumn{7}{c}{-M\bar\psi\psi}
\end{array} 
\end{eqnarray*}
\caption{Allowed representations for VLQs, with quantum numbers under $SU(2)_L$ and $U(1)_Y$ and Yukawa mixing terms in the Lagrangian. Depending on the chosen representation, the Higgs boson may be $H$ or $H^c$, therefore it has been noted as $H^{(c)}$ when necessary. The gauge invariant mass term common to all representations is a peculiar feature of VLQs.}
\label{VLrepresentations}
\end{table}
Pure mixing terms between VLQs and SM states, allowed by gauge invariance for singlets and SM-like doublet representations, have been omitted
because they can be eliminated through rotations of the states.

After the Higgs develops its VEV, VL states are allowed to mix with SM quarks: the mixing occurs in the left-handed sector for the singlet and triplet representations and in the right-handed sector for the doublet representation. The mass eigenstates will be labelled as: 
\begin{equation}
\{X_{5/3},t^\prime,b^\prime,Y_{-4/3}\}. 
\end{equation}
The mass matrices for the SM-partners $t^\prime$ and $b^\prime$ can be diagonalized by unitary $4\times4$ matrices $V^{t,b}_L$ and $V^{t,b}_R$:
\begin{eqnarray}
\left(\begin{array}{cccc} m_u & & & \\ & m_c & & \\ & & m_t & \\ & & & m_{t^\prime} \end{array}\right) &=& (V^t_L)^\dagger \cdot \mathcal{M}_t \cdot (V^t_R) \\
\left(\begin{array}{cccc} m_d & & & \\ & m_s & & \\ & & m_b & \\ & & & m_{b^\prime} \end{array}\right) &=& (V^b_L)^\dagger \cdot \mathcal{M}_b \cdot (V^b_R)
\end{eqnarray}
where the actual expressions of $\mathcal{M}_t$ and $\mathcal{M}_b$ depend on the chosen representations and on the assumptions on the mixing parameters.
The couplings with gauge bosons also depend on the chosen representations, but a common feature of every VLQ scenario is that tree-level FCNCs are developed through the mixing with SM quarks. The general form of $Zqq$ couplings with the presence of VLQs is:
\begin{eqnarray}
\setlength{\arraycolsep}{0pt}
\begin{array}{rcll}
g_{ZL}^{IJ} &=& \frac{g}{c_W}\left(T_3 - Q s_W^2\right) \delta^{IJ} & + f_L \frac{g}{c_W} (V^{t,b}_L)^{*,q^\prime I} (V^{t,b}_L)^{q^\prime J} \\
g_{ZR}^{IJ} &=& \frac{g}{c_W}\left(- Q s_W^2\right) \delta^{IJ}              & + f_R \frac{g}{c_W} (V^{t,b}_R)^{*,q^\prime I} (V^{t,b}_R)^{q^\prime J}
\end{array}
\label{Zcoupling}
\end{eqnarray}
where $I,J$ run on all quarks, including VLQs, $T_3=\pm1/2$ is the weak isospin of top or bottom in the SM, and $f_{L,R}=\{0,\pm1/2,\pm1\}$ are parameters which depend on the VLQ representation and satisfy the relation $T_3^{q^\prime}=T_3+f_L=f_R$; they are listed in Tab.~\ref{fLfR} for each representation.

\begin{table}[htb]
\begin{eqnarray*}
\begin{array}{cccccccccc}
& & \multicolumn{2}{c}{\mbox{Singlets}} & \multicolumn{3}{c}{\mbox{Doublets}} & \multicolumn{2}{c}{\mbox{Triplets}} \\
\toprule
U(1)_Y & & 2/3 & -1/3 & 7/6 & 1/6 & -5/6 & 2/3 & -1/3 \\
\midrule
t^\prime & \begin{array}{c} f_L \\ f_R \end{array} & 
\begin{array}{c} -1/2 \\  0   \end{array} & & 
\begin{array}{c} -1   \\ -1/2 \end{array} & 
\begin{array}{c}  0   \\ +1/2 \end{array} & &
\begin{array}{c} -1/2 \\  0   \end{array} & 
\begin{array}{c} +1/2 \\ +1   \end{array} \\
\midrule
b^\prime & \begin{array}{c} f_L \\ f_R \end{array} & & 
\begin{array}{c} +1/2 \\ 0    \end{array} & &
\begin{array}{c}  0   \\ -1/2 \end{array} & 
\begin{array}{c} +1   \\  1/2 \end{array} & 
\begin{array}{c} -1/2 \\ -1   \end{array} & 
\begin{array}{c} +1/2 \\  0   \end{array} 
\end{array}
\end{eqnarray*}
\caption{Neutral current parameters $f_L$ and $f_R$.}
\label{fLfR}
\end{table}

From (\ref{Zcoupling}) two consequences can be inferred:
\begin{enumerate}
\item FCNCs are present between the new state and SM quarks, but can also be induced between SM quarks themselves, if mixing with light families is allowed.
\item Even flavour conserving neutral currents ($I=J$) are modified by the presence of VLQs.
\end{enumerate}
Constraints on FCNCs coming from a large number of observations can therefore provide strong bounds on mixing parameters.

Charged currents are modified too. The general form of $Wq_1q_2$ couplings with the presence of VLQs is:
\begin{eqnarray}
\setlength{\arraycolsep}{0pt}
g_{WL}^{IJ} = \frac{g}{\sqrt{2}} (V_{CKM}^L)^{IJ} = \frac{g}{\sqrt{2}} (V^t_L)^\dagger \cdot \hat\delta_L \cdot \tilde V_{CKM}^L \cdot V^b_L \\
g_{WR}^{IJ} = \frac{g}{\sqrt{2}} (V_{CKM}^R)^{IJ} = \frac{g}{\sqrt{2}} (V^t_R)^\dagger \cdot \hat\delta_R \cdot \tilde V_{CKM}^R \cdot V^b_R
\label{VCKM}
\end{eqnarray}
where $I,J=1,2,3(,4)$ and the matrices $V^{t,b}$ may or may not be present depending on the scenario considered. The matrices $\hat\delta_{L,R}$ are defined as:
\begin{eqnarray}
\hat\delta_L=\left(\begin{array}{ccc|c} 1 & ~ & ~ & ~ \\ ~ & 1 & ~ & ~ \\ ~ & ~ & 1 & ~ \\ \hline ~ & ~ & ~ & 1 \end{array} \right) \qquad
\hat\delta_R=\left(\begin{array}{ccc|c} 0 & ~ & ~ & ~ \\ ~ & 0 & ~ & ~ \\ ~ & ~ & 0 & ~ \\ \hline ~ & ~ & ~ & 1 \end{array} \right) \qquad
\label{dLdR}
\end{eqnarray}
where the lines mean that the size of the matrices depend on the chosen scenario; in particular, $g_{WR}$ exists only if both an up- and down-type VLQ are present simultaneously. The matrices $\tilde V_{CKM}^{L,R}$ represent the misalignment between SM quarks in the left- and right-handed sector; $\tilde V_{CKM}^L$ corresponds to the measured $CKM$ matrix in the absence of VLQs. Two $CKM$ matrices can thus be defined in the presence of VLQs, for the left- and right-handed sectors. If VLQs exist, the measured $CKM$ matrix corresponds to the $3\times3$ block $(V_{CKM}^L)^{ij}$, with $i,j=1,3$. A further consequence of the introduction of VLQs is that the measured $3\times 3$ $CKM$ block is not unitary, and it is possible to check that deviations from unitarity are proportional to the mixing between SM quarks and VL states. 

Charged currents may also be present between the exotic states $\{X_{5/3},Y_{-4/3}\}$ and up- or down-type quark respectively. The couplings are:
\begin{eqnarray}
\begin{array}{rcl} 
g_W^{XI} &=& \frac{g}{\sqrt{2}} \left( (V^t_L)^{4I} + (V^t_R)^{4I} \right) \\
g_W^{YI} &=& \frac{g}{\sqrt{2}} \left( (V^b_L)^{4I} + (V^b_R)^{4I} \right) 
\end{array}
\end{eqnarray}

Finally, the couplings to the Higgs bosons can be written as:
\begin{eqnarray}
\setlength{\arraycolsep}{0pt}
\begin{array}{rcl}
C_u^{IJ} &=& \frac{1}{v} \left(\setlength{\arraycolsep}{2pt}\begin{array}{cccc} m_u & & & \\ & m_c & & \\ & & m_t & \\ & & & m_t^\prime \end{array}\right) - \frac{M}{v} (V^t_L)^{*,4I} (V^t_R)^{4J} \\[25pt]
C_d^{IJ} &=& \frac{1}{v} \left(\setlength{\arraycolsep}{2pt}\begin{array}{cccc} m_d & & & \\ & m_s & & \\ & & m_b & \\ & & & m_b^\prime \end{array}\right) - \frac{M}{v} (V^b_L)^{*,4I} (V^b_R)^{4J}
\end{array}
\end{eqnarray}
From these expressions it can be inferred that the presence of VLQ can modify the mechanism of production and decay of the Higgs boson with respect to SM predictions.

\subsection{Constraints on model parameters}

The presence of new states induces corrections to precisely measured observables of the SM both at tree level and at loop level. Tree level modifications are robust, in the sense that they can affect observables which in the SM are generated only at loop level and because they only depend on mixing parameters and new particles representations. Loop corrections are more model-dependent: however, the presence of new heavy states can result in cancellations between diagrams which can sensibly change loop-level observables. In the following a short review of the main observables which can provide constraints on the mixing parameters of VLQs is provided, considering the most recent experimental measurements; more details on analyses and formulas can be found in \cite{Branco:1986my,Lavoura:1992np,AguilarSaavedra:2002kr,Arnold:2010vs} for the singlet or SM-doublet scenarios, or in \cite{Cacciapaglia:2011fx} for the non-SM doublet $(X_{5/3}\;t^\prime)$ scenario.

\subsubsection*{Tree-level constraints}
Allowing a mixing between VLQ and SM quarks means that couplings of the type $Vq_1q_2$, where $V=W,Z$ and $q_{1,2}$ are SM quarks, receive deviations which can be observable in experimental searches. Such deviations depend on the mixing parameters of the new states and on their representations, allowing the possibility to pose strong bounds on the coupling between VLQs and SM particles. 

If the VLQs mix only with third generation SM quarks, the only affected observables are $Wtb$ and $Zbb$; since either the top or bottom quarks are mixed with the new states in all representations, the CC coupling $Wtb$ is always affected at tree level by the presence of VLQ, while the NC coupling $Zbb$ is modified at tree level only if a $b^\prime$ VLQ is present.

If the VLQs mix with lighter generations a number of observables is affected: FCNC can contribute at tree level to SM observables that otherwise would receive only loop-level contributions. The main observables which can be modified at tree-level by the presence of new VLQs are listed in the following.\\

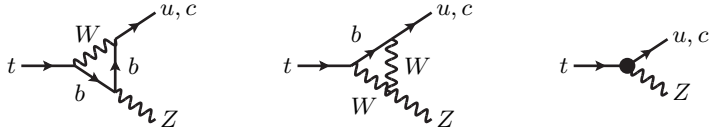
\begin{figure}
\begin{center}
 \begin{picture}(100,40)(0,-10)
  \SetWidth{1}
  \Line[arrow,arrowpos=0.5,arrowlength=2,arrowwidth=1,arrowinset=0.2](10,10)(30,10)
  \Text(8,10)[rc]{\small{$t$}}
  \Line[arrow,arrowpos=0.5,arrowlength=2,arrowwidth=1,arrowinset=0.2](30,10)(45,0)
  \Text(30,0)[lc]{\small{$b$}}
  \Line[arrow,arrowpos=0.5,arrowlength=2,arrowwidth=1,arrowinset=0.2](45,0)(45,20)
  \Text(50,10)[lc]{\small{$b$}}
  \Line[arrow,arrowpos=0.5,arrowlength=2,arrowwidth=1,arrowinset=0.2](45,20)(60,30)
  \Text(62,30)[lc]{\small{$u,c$}}
  \Photon(30,10)(45,20){2}{4}
  \Text(30,22)[lc]{\small{$W$}}
  \Photon(45,0)(60,-10){2}{4}
  \Text(62,-10)[lc]{\small{$Z$}}
 \end{picture}
 \begin{picture}(100,40)(0,-10)
  \SetWidth{1}
  \Line[arrow,arrowpos=0.5,arrowlength=2,arrowwidth=1,arrowinset=0.2](10,10)(30,10)
  \Text(8,10)[rc]{\small{$t$}}
  \Line[arrow,arrowpos=0.5,arrowlength=2,arrowwidth=1,arrowinset=0.2](30,10)(45,20)
  \Text(30,22)[lc]{\small{$b$}}
  \Photon(30,10)(45,0){2}{4}
  \Text(30,-5)[lc]{\small{$W$}}
  \Photon(45,0)(45,20){2}{4}
  \Text(50,10)[lc]{\small{$W$}}
  \Line[arrow,arrowpos=0.5,arrowlength=2,arrowwidth=1,arrowinset=0.2](45,20)(60,30)
  \Text(62,30)[lc]{\small{$u,c$}}
  \Photon(45,0)(60,-10){2}{4}
  \Text(62,-10)[lc]{\small{$Z$}}
 \end{picture}
 \begin{picture}(100,40)(0,-10)
  \SetWidth{1}
  \Line[arrow,arrowpos=0.5,arrowlength=2,arrowwidth=1,arrowinset=0.2](10,10)(30,10)
  \Text(8,10)[rc]{\small{$t$}}
  \Line[arrow,arrowpos=0.5,arrowlength=2,arrowwidth=1,arrowinset=0.2](30,10)(45,20)
  \Text(47,20)[lc]{\small{$u,c$}}
  \Photon(30,10)(45,0){2}{4}
  \Text(47,0)[lc]{\small{$Z$}}
  \Vertex(30,10){3}
 \end{picture}
\end{center}
\caption{Feynman diagrams for FCNC rare top decays in the SM and induced at tree level by the presence of VLQs.}
\label{FCNCtdecay}
\end{figure}

\noindent{\it Rare FCNC top decays}\\
FCNC processes describing top decays in the SM and in a VLQ scenario are represented in Fig.~\ref{FCNCtdecay}. 
Current experimental bounds are set by the CMS experiment, which has performed a study of rare top decay with an integrated luminosity of $5.0~\mathrm{fb}^{-1}$ \cite{Chatrchyan:2012sd}, finding a limit $BR(t\to Zq)<0.24\%$, whereas the SM prediction is $BR(t\to Zq)=\mathcal{O}(10^{-14})$. This limit can be translated into bounds on combinations of non-diagonal mixing matrix elements in both the left- and right-handed sector.\\

\noindent{\it $Zqq$ couplings}\\
As already stated, flavour conserving couplings are also affected by the presence of new states. The left- and right-handed couplings $Zqq$ have been measured at LEP1 \cite{ALEPH:2005ab} and the strongest bounds come from the charm and bottom quarks:
\begin{eqnarray}
\setlength{\arraycolsep}{0pt}
\left\{\begin{array}{l} g_{ZL}^c = 0.3453\pm0.0036 \\ g_{ZR}^c = -0.1580\pm0.0051 \end{array}\right. \qquad 
\left\{\begin{array}{l} g_{ZL}^b = -0.4182\pm0.00315 \\ g_{ZR}^b = 0.0962\pm0.0063 \end{array}\right.
\end{eqnarray}
Deviations from SM predictions can also be parametrized through the quantities:
\begin{eqnarray}
R_q = {\Gamma(Z\to q \bar q) \over \Gamma(Z\to {\rm hadrons})} = R_q^{\rm SM} (1 + \delta R_q)\\ 
A_q = {(g_{ZL}^q)^2 - (g_{ZR}^q)^2 \over (g_{ZL}^q)^2 + (g_{ZR}^q)^2)} = A_q^{\rm SM} (1 +\delta A_q)
\end{eqnarray}
with $q=c,b$ and $R_q^{\rm SM}$ and $A_q^{\rm SM}$ are the SM predictions. The observed and predicted values are \cite{Baak:2012kk}: 
\begin{eqnarray}
\setlength{\arraycolsep}{0pt}
\begin{array}{ll}
\left\{\begin{array}{l} R_c^{\rm exp} = 0.1721  \pm 0.0030  \\ R_c^{\rm SM} = 0.17223 \pm 0.00006 \end{array}\right. & 
\quad\left\{\begin{array}{l} R_b^{\rm exp} = 0.21629 \pm 0.00066 \\ R_b^{\rm SM} = 0.21474 \pm 0.00003 \end{array}\right. \\[10pt]
\left\{\begin{array}{l} A_c^{\rm exp} = 0.670 \pm 0.027 \\ A_c^{\rm SM} = 0.6680^{+0.00025}_{-0.00038} \end{array}\right. & 
\quad\left\{\begin{array}{l} A_b^{\rm exp} = 0.923 \pm 0.020 \\ A_b^{\rm SM} = 0.93464^{+0.00004}_{-0.00007} \end{array}\right. 
\end{array}
\end{eqnarray}

Bounds on matrix elements $(V^{t,b}_{L,R})^{4i}$ can be obtained evaluating the contributions induced by VLQs to the measured quantities. The contributions are at tree level only if there is mixing between the VLQ and the considered quark. The contribution to the coupling is at loop level if there is no mixing (e.g. the contribution of a $t^\prime$ singlet to $Zbb$).\\

\begin{figure}[htb]
\begin{center}
 \begin{picture}(150,40)(0,-10)
  \SetWidth{1}
  \Text(20,10)[rc]{$D^0\bigg\{$}
  \Line[arrow,arrowpos=0.5,arrowlength=2,arrowwidth=1,arrowinset=0.2](30,20)(50,20)
  \Text(28,20)[rc]{\small{$c$}}
  \Line[arrow,arrowpos=0.5,arrowlength=2,arrowwidth=1,arrowinset=0.2](50,20)(50,0)
  \Text(46,10)[rc]{\small{$d$}}
  \Line[arrow,arrowpos=0.5,arrowlength=2,arrowwidth=1,arrowinset=0.2](50,0)(30,0)
  \Text(28,0)[rc]{\small{$u$}}
  \Photon(50,20)(70,20){2}{4}
  \Text(60,25)[cb]{\small{$W$}}
  \Photon(50,0)(70,0){2}{4}
  \Text(60,-5)[ct]{\small{$W$}}
  \Line[arrow,arrowpos=0.5,arrowlength=2,arrowwidth=1,arrowinset=0.2](70,20)(90,20)
  \Text(92,20)[lc]{\small{$u$}}
  \Line[arrow,arrowpos=0.5,arrowlength=2,arrowwidth=1,arrowinset=0.2](70,0)(70,20)
  \Text(74,10)[lc]{\small{$d$}}
  \Line[arrow,arrowpos=0.5,arrowlength=2,arrowwidth=1,arrowinset=0.2](90,0)(70,0)
  \Text(92,0)[lc]{\small{$c$}}
  \Text(100,10)[lc]{$\bigg\}\bar D^0$}
 \end{picture}
 \begin{picture}(130,40)(0,-10)
  \SetWidth{1}
  \Text(20,10)[rc]{$D^0\bigg\{$}
  \Line[arrow,arrowpos=0.5,arrowlength=2,arrowwidth=1,arrowinset=0.2](30,20)(50,20)
  \Text(28,20)[rc]{\small{$c$}}
  \Line[arrow,arrowpos=0.5,arrowlength=2,arrowwidth=1,arrowinset=0.2](50,20)(50,0)
  \Text(46,10)[rc]{\small{$d$}}
  \Line[arrow,arrowpos=0.5,arrowlength=2,arrowwidth=1,arrowinset=0.2](50,0)(30,0)
  \Text(28,0)[rc]{\small{$u$}}
  \Photon(50,20)(70,20){2}{4}
  \Text(60,25)[cb]{\small{$W$}}
  \Photon(50,0)(70,0){2}{4}
  \Text(60,-5)[ct]{\small{$W$}}
  \Line[arrow,arrowpos=0.5,arrowlength=2,arrowwidth=1,arrowinset=0.2](70,20)(90,20)
  \Text(92,20)[lc]{\small{$l^+$}}
  \Line[arrow,arrowpos=0.5,arrowlength=2,arrowwidth=1,arrowinset=0.2](70,0)(70,20)
  \Text(74,10)[lc]{\small{$\nu_i$}}
  \Line[arrow,arrowpos=0.5,arrowlength=2,arrowwidth=1,arrowinset=0.2](90,0)(70,0)
  \Text(92,0)[lc]{\small{$l^-$}}
 \end{picture}
 \begin{picture}(150,40)(0,-10)
  \SetWidth{1}
  \Text(20,10)[rc]{$D^0\bigg\{$}
  \Line[arrow,arrowpos=0.5,arrowlength=2,arrowwidth=1,arrowinset=0.2](30,20)(50,20)
  \Text(28,20)[rc]{\small{$c$}}
  \Photon(50,20)(50,0){2}{4}
  \Text(46,10)[rc]{\small{$W$}}
  \Line[arrow,arrowpos=0.5,arrowlength=2,arrowwidth=1,arrowinset=0.2](50,0)(30,0)
  \Text(28,0)[rc]{\small{$u$}}
  \Line[arrow,arrowpos=0.5,arrowlength=2,arrowwidth=1,arrowinset=0.2](50,20)(70,10)
  \Text(65,18)[cb]{\small{$d_i$}}
  \Line[arrow,arrowpos=0.5,arrowlength=2,arrowwidth=1,arrowinset=0.2](70,10)(50,0)
  \Text(65,2)[ct]{\small{$d_i$}}
  \Photon(70,10)(90,10){2}{4}
  \Text(80,15)[cb]{\small{$\gamma,Z$}}
  \Line[arrow,arrowpos=0.5,arrowlength=2,arrowwidth=1,arrowinset=0.2](90,10)(110,20)
  \Text(112,20)[lc]{\small{$l^+$}}
  \Line[arrow,arrowpos=0.5,arrowlength=2,arrowwidth=1,arrowinset=0.2](110,0)(90,10)
  \Text(112,0)[lc]{\small{$l^-$}}
 \end{picture}\\

 \begin{picture}(150,40)(0,-10)
  \SetWidth{1}
  \Text(20,10)[rc]{$D^0\bigg\{$}
  \Line[arrow,arrowpos=0.5,arrowlength=2,arrowwidth=1,arrowinset=0.2](30,20)(50,10)
  \Text(28,20)[rc]{\small{$c$}}
  \Line[arrow,arrowpos=0.5,arrowlength=2,arrowwidth=1,arrowinset=0.2](50,10)(30,0)
  \Text(28,0)[rc]{\small{$u$}}
  \Photon(50,10)(70,10){2}{4}
  \Text(60,15)[cb]{\small{$Z$}}
  \Line[arrow,arrowpos=0.5,arrowlength=2,arrowwidth=1,arrowinset=0.2](70,10)(90,20)
  \Text(92,20)[lc]{\small{$u$}}
  \Line[arrow,arrowpos=0.5,arrowlength=2,arrowwidth=1,arrowinset=0.2](90,0)(70,10)
  \Text(92,0)[lc]{\small{$c$}}
  \Text(100,10)[lc]{$\bigg\}\bar D^0$}
  \Vertex(50,10){3}
  \Vertex(70,10){3}
 \end{picture}
 \begin{picture}(150,40)(0,-10)
  \SetWidth{1}
  \Text(20,10)[rc]{$D^0\bigg\{$}
  \Line[arrow,arrowpos=0.5,arrowlength=2,arrowwidth=1,arrowinset=0.2](30,20)(50,10)
  \Text(28,20)[rc]{\small{$c$}}
  \Line[arrow,arrowpos=0.5,arrowlength=2,arrowwidth=1,arrowinset=0.2](50,10)(30,0)
  \Text(28,0)[rc]{\small{$u$}}
  \Photon(50,10)(70,10){2}{4}
  \Text(60,15)[cb]{\small{$Z$}}
  \Line[arrow,arrowpos=0.5,arrowlength=2,arrowwidth=1,arrowinset=0.2](70,10)(90,20)
  \Text(92,20)[lc]{\small{$l^+$}}
  \Line[arrow,arrowpos=0.5,arrowlength=2,arrowwidth=1,arrowinset=0.2](90,0)(70,10)
  \Text(92,0)[lc]{\small{$l^-$}}
  \Vertex(50,10){3}
 \end{picture}
\end{center}
\caption{Feynman diagrams for $D^0$ mixing and decay at loop level in the SM (first row) and at tree level with the presence of VLQs (second row).}
\label{D0mixinganddecay}
\end{figure}
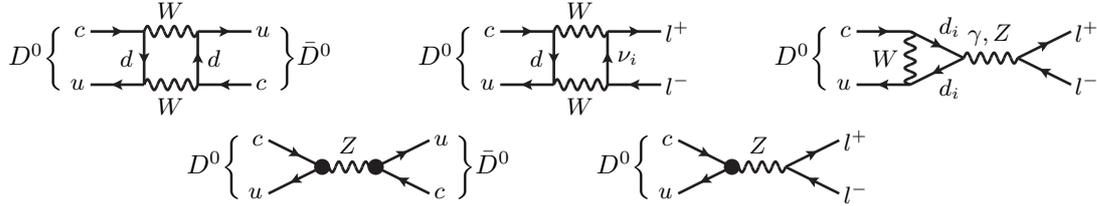

\noindent{\it Meson mixing and decays}\\
FCNCs induced by the presence of VLQs can play a relevant role in meson mixing and decay. Some processes which in the SM can only occur at loop level may be generated at tree-level through FCNCs. Mixing parameters and branching ratios for a large number of mesons have been measured accurately, providing strong constraints on VLQ mixing parameters. The Feynman diagrams of $D^0$ mixing and decay in the SM and with the contribution of VLQs are shown in Fig.~\ref{D0mixinganddecay}.

Meson mixing and decays have been widely studied in literature, and an analytical description of the contributions of VLQs for specific processes is beyond the purposes of this analysis. Detailed studies for specific scenarios can be found in \cite{Barenboim:2001fd,Lee:2004me,Cacciapaglia:2011fx,Berger:2012ec}. 
It can be noticed, however, that observables in the down sector have been measured with much more precision, with respect to the up sector: the contribution of down-type VLQs with mixing to light generation can be therefore strongly constrained by bounds coming from the flavour sector.\\

\noindent{\it Atomic parity violation}\\
A strong bound on mixing parameters between VLQs and the first quark generation comes from measurements of the atomic parity violation, which provides information about $Zuu$ and $Zdd$ couplings. The weak charge of a nucleus is defined as \cite{Deandrea:1997wk}:
\begin{equation}
 Q_W = {2 c_W \over g} \left[(2Z+N)(g^u_{ZL} + g^u_{ZR}) + (Z+2N) (g^d_{ZL}+g^d_{ZR})\right]
\end{equation}
where $g^{u,d}_{ZL,ZR}$ are the left- and right-handed couplings of light quarks with the $Z^0$ boson and $Z$ and $N$ are the number of protons and neutrons in the nucleus, respectively. The most precise test on atomic parity violation are in Cesium $^{133}$Cs and Tallium $^{204}$ Tl \cite{Nakamura:2010zzi}:
\begin{eqnarray}
\setlength{\arraycolsep}{0pt}
\left\{\begin{array}{l}
\left. Q_W (\mbox{$^{133}$Cs})\right|_{\rm exp.} = -73.20\pm 0.35 \\
\left. Q_W (\mbox{$^{133}$Cs})\right|_{\rm SM} = -73.15 \pm 0.02  
\end{array}\right.\hskip 30pt
\left\{\begin{array}{l}
\left. Q_W (\mbox{$^{204}$Tl})\right|_{\rm exp.} = -116.4\pm 3.6  \\
\left. Q_W (\mbox{$^{204}$Tl})\right|_{\rm SM} = -116.76 \pm 0.04
\end{array}\right.
\end{eqnarray}
Deviations on flavour conserving neutral couplings given by the contribution of VLQs, defined in (\ref{Zcoupling}), provide bounds on a combination of the matrix elements $(V^{t,b}_{L,R})^{41}$:
\begin{equation}
\delta Q_W^{VL} = 2\left[(2Z+N)\left((T_3^{t^\prime}-{1\over2})|(V_L^t)^{41}|^2+T_3^{t^\prime}|(V_R^t)^{41}|^2\right)+(Z+2N)\left((T_3^{b^\prime}+{1\over2})|(V_L^b)^{41}|^2+T_3^{b^\prime}|(V_R^b)^{41}|^2\right)\right] 
\end{equation}
\\

\noindent{\it CKM matrix}\\
CKM matrix entries have been measured very precisely both at tree and loop level. Deviations from the standard $3\times 3$ unitary $V_{CKM}$ have also been studied extensively in the context of new physics scenarios. The contribution of VLQs strongly depend on the scenario considered. The very presence of a $CKM$ matrix in the right-handed sector is linked to the existence of both a top and a bottom VL partner, as evident from (\ref{dLdR}). The $CKM$ matrix can also contain new phases, which can induce CP violations.

\subsubsection*{Loop-level constraints}
Loop constraints are more model dependent: deviations from SM predictions may occur only if specific particles circulate in loops, but the particle content of the theory depends on which representation the VLQs belong to, and in many cases, SM quantities are not affected at all. The main observables which can be affected by VLQs at loop level are shortly described in the following.\\

\noindent{\it EW precision tests}\\
Regardless of the representation the VLQ belongs to, the new states induce modification at loop level to the va\-cu\-um polarizations of electroweak gauge bosons, which are parametrised by the oblique parameters $S,T,U$ \cite{Peskin:1991sw}. The impact of VLQs in the case of mixing only with third generation has been first analysed in \cite{Lavoura:1992np}, while a more recent analysis for all VLQ representations mixing with third generation is in \cite{Cacciapaglia:2010vn}. In \cite{Cynolter:2008ea} a study for a specific scenario with the presence of both a singlet and a doublet is presented, and an analysis of the $S,T,U$ parameters for the singlet and SM-doublet representations taking into account variations of the Higgs mass and emphasizing the decoupling properties of VLQs is in \cite{Dawson:2012di}. \\

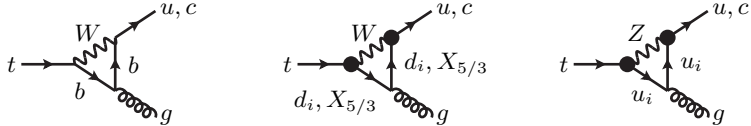
\begin{figure}[htb]
\begin{center}
 \begin{picture}(100,40)(0,-10)
  \SetWidth{1}
  \Line[arrow,arrowpos=0.5,arrowlength=2,arrowwidth=1,arrowinset=0.2](10,10)(30,10)
  \Text(8,10)[rc]{\small{$t$}}
  \Line[arrow,arrowpos=0.5,arrowlength=2,arrowwidth=1,arrowinset=0.2](30,10)(45,0)
  \Text(30,0)[lc]{\small{$b$}}
  \Line[arrow,arrowpos=0.5,arrowlength=2,arrowwidth=1,arrowinset=0.2](45,0)(45,20)
  \Text(50,10)[lc]{\small{$b$}}
  \Line[arrow,arrowpos=0.5,arrowlength=2,arrowwidth=1,arrowinset=0.2](45,20)(60,30)
  \Text(62,30)[lc]{\small{$u,c$}}
  \Photon(30,10)(45,20){2}{4}
  \Text(30,22)[lc]{\small{$W$}}
  \Gluon(45,0)(60,-10){2}{4}
  \Text(62,-10)[lc]{\small{$g$}}
 \end{picture}
 \begin{picture}(100,40)(0,-10)
  \SetWidth{1}
  \Line[arrow,arrowpos=0.5,arrowlength=2,arrowwidth=1,arrowinset=0.2](10,10)(30,10)
  \Text(8,10)[rc]{\small{$t$}}
  \Vertex(30,10){3}
  \Line[arrow,arrowpos=0.5,arrowlength=2,arrowwidth=1,arrowinset=0.2](30,10)(45,0)
  \Text(40,0)[rt]{\small{$d_i,X_{5/3}$}}
  \Line[arrow,arrowpos=0.5,arrowlength=2,arrowwidth=1,arrowinset=0.2](45,0)(45,20)
  \Text(50,10)[lc]{\small{$d_i,X_{5/3}$}}
  \Vertex(45,20){3}
  \Line[arrow,arrowpos=0.5,arrowlength=2,arrowwidth=1,arrowinset=0.2](45,20)(60,30)
  \Text(62,30)[lc]{\small{$u,c$}}
  \Photon(30,10)(45,20){2}{4}
  \Text(30,22)[lc]{\small{$W$}}
  \Gluon(45,0)(60,-10){2}{4}
  \Text(62,-10)[lc]{\small{$g$}}
 \end{picture}
 \begin{picture}(100,40)(0,-10)
  \SetWidth{1}
  \Line[arrow,arrowpos=0.5,arrowlength=2,arrowwidth=1,arrowinset=0.2](10,10)(30,10)
  \Text(8,10)[rc]{\small{$t$}}
  \Vertex(30,10){3}
  \Line[arrow,arrowpos=0.5,arrowlength=2,arrowwidth=1,arrowinset=0.2](30,10)(45,0)
  \Text(40,0)[rt]{\small{$u_i$}}
  \Line[arrow,arrowpos=0.5,arrowlength=2,arrowwidth=1,arrowinset=0.2](45,0)(45,20)
  \Text(50,10)[lc]{\small{$u_i$}}
  \Vertex(45,20){3}
  \Line[arrow,arrowpos=0.5,arrowlength=2,arrowwidth=1,arrowinset=0.2](45,20)(60,30)
  \Text(62,30)[lc]{\small{$u,c$}}
  \Photon(30,10)(45,20){2}{4}
  \Text(30,22)[lc]{\small{$Z$}}
  \Gluon(45,0)(60,-10){2}{4}
  \Text(62,-10)[lc]{\small{$g$}}
 \end{picture}
\end{center}
\caption{Feynman diagrams for FCNC top decays into $gq$ in the SM (left) and induced at loop level by the presence of VLQs (centre and right). In the diagrams, $u_i=u,c,t,t^\prime$ and $d_i=d,s,b,b^\prime$.}
\label{toptogq}
\end{figure}

\noindent{\it Rare top decay}\\
VLQs may contribute at loop level to FCNC top decays which are GIM suppressed in the SM, as the decay $t\to gq$ shown in Fig.~\ref{toptogq}.
The presence of modified couplings can be competitive with the SM diagram. The strongest bounds on $t\to gq$ decay have been set by ATLAS \cite{Aad:2012gd} using a sample with integrated luminosity of 2.05 fb$^{-1}$. The obtained branching ratios are:
\begin{eqnarray}
\setlength{\arraycolsep}{0pt}
\left\{\begin{array}{l}
BR(t\to gu) < 5.7 \times 10^{-5} \\
BR(t\to gc) < 2.7 \times 10^{-4}
\end{array}\right.
\end{eqnarray}
whereas the SM prediction is $BR(t\to gq)=\mathcal{O}(10^{-14})$.

\begin{figure}[htb]
\begin{center}
 \begin{picture}(160,40)(0,-10)
  \SetWidth{1}
  \Text(20,10)[rc]{$K^0\bigg\{$}
  \Line[arrow,arrowpos=0.5,arrowlength=2,arrowwidth=1,arrowinset=0.2](30,20)(70,20)
  \Text(28,20)[rc]{\small{$d$}}
  \Line[arrow,arrowpos=0.5,arrowlength=2,arrowwidth=1,arrowinset=0.2](70,20)(70,0)
  \Text(66,10)[rc]{\small{$u,c,t,t^\prime$}}
  \Line[arrow,arrowpos=0.5,arrowlength=2,arrowwidth=1,arrowinset=0.2](70,0)(30,0)
  \Text(28,0)[rc]{\small{$s$}}
  \Photon(70,20)(90,20){2}{4}
  \Text(80,25)[cb]{\small{$W$}}
  \Photon(70,0)(90,0){2}{4}
  \Text(80,-5)[ct]{\small{$W$}}
  \Line[arrow,arrowpos=0.5,arrowlength=2,arrowwidth=1,arrowinset=0.2](90,20)(130,20)
  \Text(132,20)[lc]{\small{$s$}}
  \Line[arrow,arrowpos=0.5,arrowlength=2,arrowwidth=1,arrowinset=0.2](90,0)(90,20)
  \Text(94,10)[lc]{\small{$u,c,t,t^\prime$}}
  \Line[arrow,arrowpos=0.5,arrowlength=2,arrowwidth=1,arrowinset=0.2](130,0)(90,0)
  \Text(132,0)[lc]{\small{$d$}}
  \Text(140,10)[lc]{$\bigg\}\bar K^0$}
 \end{picture}\hskip 10pt
 \begin{picture}(160,40)(0,-10)
  \SetWidth{1}
  \Text(20,10)[rc]{$K^0\bigg\{$}
  \Line[arrow,arrowpos=0.5,arrowlength=2,arrowwidth=1,arrowinset=0.2](30,20)(50,20)
  \Text(28,20)[rc]{\small{$d$}}
  \Line[arrow,arrowpos=0.5,arrowlength=2,arrowwidth=1,arrowinset=0.2](50,0)(30,0)
  \Text(28,0)[rc]{\small{$s$}}
  \Photon(50,20)(50,0){2}{4}
  \Text(46,10)[rc]{\small{$W$}}
  \Line[arrow,arrowpos=0.5,arrowlength=2,arrowwidth=1,arrowinset=0.2](50,20)(90,20)
  \Text(70,23)[cb]{\small{$u,c,t,t^\prime$}}
  \Line[arrow,arrowpos=0.5,arrowlength=2,arrowwidth=1,arrowinset=0.2](90,0)(50,0)
  \Text(70,-3)[ct]{\small{$u,c,t,t^\prime$}}
  \Photon(90,0)(90,20){2}{4}
  \Text(94,10)[lc]{\small{$W$}}
  \Line[arrow,arrowpos=0.5,arrowlength=2,arrowwidth=1,arrowinset=0.2](90,20)(110,20)
  \Text(112,20)[lc]{\small{$s$}}
  \Line[arrow,arrowpos=0.5,arrowlength=2,arrowwidth=1,arrowinset=0.2](110,0)(90,0)
  \Text(112,0)[lc]{\small{$d$}}
  \Text(120,10)[lc]{$\bigg\}\bar K^0$}
 \end{picture}
\end{center}
\caption{Feynman diagrams for $K^0$ mixing with the presence of a VLQ $t^\prime$ in the loops.}
\label{K0mixing}
\end{figure}
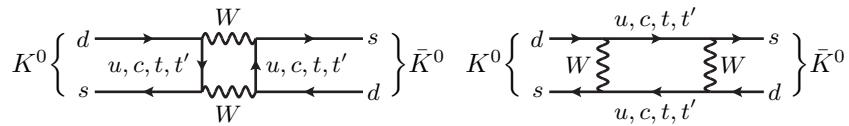

\noindent{\it Flavour physics}\\
VLQs can also play a role in flavour physics: new states can circulate in loops together with SM particles, and even small corrections can spoil cancellations within loop diagrams, producing observable effects. Such phenomena can be particularly relevant for meson mixing, especially if VLQs belong to a representation for which tree level diagrams such those of Fig. ~\ref{D0mixinganddecay} are not allowed. A $t^\prime$ can contribute to the mixing of mesons in the down-sector, as in Fig. ~\ref{K0mixing}, and if it belongs to a singlet or non-SM doublet representation, this is the only contribution of the new state to the mixing. Analogous considerations can be done for the other VLQs.
 
VLQ contributions to meson mixing at loop level are strongly model dependent, a general analysis of their contribution is beyond the aims of this study. Detailed computations in specific models are present in literature, though. In \cite{Picek:2008dd} the role of a VL top singlet, mixing dominantly with the top quark, in the rare decays $K \to \pi\nu\bar\nu$, $B\to\pi(K)\nu\bar\nu$ and $B_{s,d}\to\mu^+\mu^-$ is considered, finding corrections from 20\% for $K$ and $B$ rare decays, up to 50\% for $B_{s,d}$ decays. Analogous conclusions have been obtained in \cite{Cacciapaglia:2011fx}, where corrections to the $\Delta F=2$ mixings in the $K^0$ and $B^0$ sectors given by the presence of a $t^\prime$ in the non-SM doublet with mixing to all quark families have been considered, finding sizable contributions, up to 60\%, only for the phase of the $B_s^0 - \bar B_s^0$ mixing.

\section{Past and Current Searches}
\label{Sect:searches}

Various searches of new heavy states have been undertaken both at Tevatron and at the LHC, though no evidence for the existence of other quarks, beside those of the SM, has been obtained. Direct bounds on heavy chiral quarks can be interpreted as bound on VLQs, but it must be stressed that decay channels of VLQs are different from decay channels of heavy chiral quarks. For VLQs charged and neutral currents can have similar branching ratios, therefore searches performed with specific assumptions on the heavy state decay channel can give a rough idea of the bounds on VLQ mass, once rescaled with the actual branching ratios in the specific channel. In the following sections an overview will be given of all available searches of heavy quarks at Tevatron and at the LHC, both chiral and vector-like, focusing on the assumptions that have been made to obtain the bounds on the heavy quark masses. Searches of chiral quarks have been included only for reference purpose: a reinterpretation of bounds considering VLQ 
branching ratios is beyond the scope of this review, but limits coming from searches of chiral quarks can only apply to VLQ scenarios only after appropriate rescaling. More details on single searches (kinematic cuts, detector parameters \dots) can be found in the original publications.

\subsection{Tevatron}

The studies refer to the Tevatron RunII with a center of mass energy of 1.96 TeV. The bounds are presented for each experiment with increasing integrated luminosity.


%

\subsubsection*{CDF @ $1.06~\mathrm{fb}^{-1}$}

In \cite{Aaltonen:2007je}, a search for $b^\prime$ pair production is performed. The $b^\prime$ is expected to decay 100\% in $Zb$ and one $Z$ is then required to decay leptonically. The analysis select events with one reconstructed $Z$ and $\geq$3 jets. The obtained bound is:
\begin{equation}
 m_{b^\prime} > 268~\mathrm{GeV}~~\mbox{at 95\% C.L.} \qquad \mathrm{BR}(b^\prime\to Zb)=100\%
\end{equation}

\subsubsection*{CDF @ $2.7~\mathrm{fb}^{-1}$}

In \cite{Aaltonen:2009nr}, a search for pair production of a heavy quark $Q$ which promptly decays 100\% into $Wt$ is performed. The signal considered is $l^\pm l^\pm b j E_T\hspace*{-14pt}\diagup$. The results have been interpreted in terms of bounds on $b^\prime$ mass or on the mass of an exotic $T_{5/3}$ state (this bound implies the existence of a further heavy $B$ quark):
\begin{equation}
 m_{b^\prime} , m_{B} > 338~\mathrm{GeV}~~\mbox{or}~~m_{T_{5/3}} > 365~\mathrm{GeV}~~\mbox{at 95\% C.L.} \qquad \mathrm{BR}(Q\to W^\pm t)=100\%
\end{equation}

%

%

\subsubsection*{CDF @ $4.8~\mathrm{fb}^{-1}$}
For this integrated luminosity two analyses are available.

In \cite{Aaltonen:2011vr}, a search for pair production of $b^\prime$ decaying promptly to $Wt$ is performed. One $W$ is then required to decay leptonically and the analysis looks for events with one lepton ($e$ or $\mu$), $\geq$5 jets, of which at least one b-jet, and missing transverse energy. The obtained bound is:
\begin{equation}
 m_{b^\prime} > 372~\mathrm{GeV}~~\mbox{at 95\% C.L.} \qquad \mathrm{BR}(b^\prime\to Wt)=100\%
\end{equation}

In \cite{Aaltonen:2011rr}, a search for pair production of a heavy particle $T^\prime$ decaying to $tX$ where X is an invisible (DM) particle is performed. The analysis considers the lepton+jets channel, requiring one of the $W$s coming from the tops to decay leptonically. The search is analogous to top pair production, but for the presence of missing transverse momentum. Bounds are given for the combination of $T^\prime$ and $X$ masses, and in particular, the analysis excludes the presence of a $T^\prime$ with $m_{T^\prime}\leq360$ GeV for $m_X\leq100$ GeV.

\subsubsection*{CDF @ $5.6~\mathrm{fb}^{-1}$}

In \cite{Aaltonen:2011na}, a search for $t^\prime$ pair production is performed. The $t^\prime$ is expected to decay 100\% in $Wq$, where $q$ can be a light quark or a bottom quark, and one $W$ is then required to decay leptonically. The signal under consideration is therefore $t^\prime \bar t^\prime \to WqWq \to l\nu qqqq$. This analysis considers events with one high-$p_T$ lepton ($e$ or $\mu$), large missing tranverse energy and $\geq$4 jets; for the $t\to Wb$ search at least one b-jet must be identified. The obtained bound is:
\begin{eqnarray}
 m_{t^\prime} > 358~\mathrm{GeV}~~\mbox{at 95\% C.L.} \qquad \mathrm{BR}(t^\prime\to Wb)=100\% \\
 m_{t^\prime} > 340~\mathrm{GeV}~~\mbox{at 95\% C.L.} \qquad \mathrm{BR}(t^\prime\to Wq)=100\% 
\end{eqnarray}

\subsubsection*{CDF @ $5.7~\mathrm{fb}^{-1}$}
For this integrated luminosity two analyses are available.

In \cite{Aaltonen:2011tq}, a search for pair production of a heavy particle $T^\prime$ decaying to $tX$ where X is an invisible (DM) particle is performed. The search considers events in the full hadronic channel, requiring 5$\leq$ jets $\leq$10 and missing transverse energy. Bounds are given for the combination of $T^\prime$ and $X$ masses, and in particular, the analysis excludes the presence of a $T^\prime$ with $m_{T^\prime}\leq400$ GeV for $m_X\leq70$ GeV.

In \cite{CDF10261}, a search for single production of heavy quarks is performed. The heavy quarks are expected to decay 100\% in $Wq$, where q is a SM quark of the first generation. The signal event have the topology $W+2j$, where the $W$ is required to decay leptonically. Due to the fact that single production is model dependent, the bounds on the cross section and couplings of the heavy quarks with SM quarks are given for different masses of the heavy quarks, ranging from $300$ GeV to $600$ GeV.


%

\subsubsection*{D0 @ $5.3~\mathrm{fb}^{-1}$}

In \cite{Abazov:2011vy} a search for $t^\prime$ pair production is performed. The $t^\prime$ is supposed to decay 100\% to $Wq$, where q is a light quark. The analysis selects final states with one isolated lepton, $\geq4$ jets and missing transverse energy, corresponding to one $W$ decaying leptonically and the other hadronically. Combining the $e$+jets and $\mu$+jets channels, the obtained bound is:
\begin{equation}
 m_{t^\prime} > 285~\mathrm{GeV}~~\mbox{at 95\% C.L.} \qquad \mathrm{BR}(t^\prime\to Wq)=100\%
\end{equation}

\subsubsection*{D0 @ $5.4~\mathrm{fb}^{-1}$}

In \cite{Abazov:2010ku} a search for single production of VLQs is performed. A specific model is considered \cite{Atre:2008iu}, in which there are two degenerate VL doublets, with hypercharge 7/6 and 1/6, which interact only with the first generation of SM quarks. The analysis searches for final states with either a $W$ or $Z$ boson and two jets. One jet comes from the VLQ decay and the other is produced in asociation with the VLQ. The gauge boson is then required to decay leptonically ($e$ or $\mu$) and events with exactly one lepton (case of $W$) or exactly two leptons (case of $Z$) and $\geq$2 jets are selected. Results are given for different choices of the coupling parameters and assumptions on BRs:
\begin{eqnarray}
\left.\begin{array}{c} m_{b^\prime} > 693~\mathrm{GeV}~~\mbox{at 95\% C.L.} \qquad \mathrm{BR}(b^\prime\to Wq)=100\% \\
m_{t^\prime} > 551~\mathrm{GeV}~~\mbox{at 95\% C.L.} \qquad \mathrm{BR}(t^\prime\to Zq)=100\% \end{array} \right\} & \mbox{no coupling with down quark}\\
\left.\begin{array}{c} m_{t^\prime} > 403~\mathrm{GeV}~~\mbox{at 95\% C.L.} \qquad \mathrm{BR}(t^\prime\to Wq)=100\% \\
m_{b^\prime} > 430~\mathrm{GeV}~~\mbox{at 95\% C.L.} \qquad \mathrm{BR}(b^\prime\to Zq)=100\% \end{array} \right\} & \mbox{no coupling with up quark}
\end{eqnarray}

\subsection{LHC}

The studies have been performed for a center of mass energy of 7 TeV. The bounds are presented for each experiment with increasing integrated luminosity.


%

\subsubsection*{ATLAS @ $1.04~\mathrm{fb}^{-1}$}
For this integrated luminosity several analyses are available.

In \cite{Aad:2011wc}, a search for pair production of a heavy particle $T$ decaying to $tA_0$ where $A_0$ is an invisible (DM) particle is performed. The search is performed in the $t\bar t+ MET$ channel, where one of the $W$s coming from tops decays leptonically and the other hadronically. The final state analysed contains one isolated lepton with large $p_T$, $\geq$4 jets and missing transverse energy. Bounds are given for combinations of $T$ and $A_0$ masses, excluding heavy quarks up to $M_T=420$ GeV for stable neutral states up to $M_{A_0}=140$ GeV.


In \cite{Aad:2012xc} a search for pair production of $t^\prime$ with subsequent decay into $Wb$ is performed. One of the $W$s is then required to decay leptonically. This search is conducted on final states containing one isolated lepton ($e$ or $\mu$) with high transverse momentum, high missing transverse momentum an $\geq$3 jets. The obtained bound is:
\begin{equation}
 m_{t^\prime} > 404~\mathrm{GeV}~~\mbox{at 95\% C.L.} \quad \mathrm{BR}(t^\prime\to Wb)=100\%
\end{equation}

In \cite{Aad:2012bt} a search for pair production of a fourth generation quark $Q$ decaying 100\% in $Wq$, where $q=u,d,c,s,b$ and with subsequent decay of both W into leptons. The analysis considers a final state with 2 oppsite-sign leptons, $\geq$2 jets and missing energy. The obtained bound on the new state, applicable to all heavy quarks decaying to $Wq$ is:
\begin{equation}
 m_{Q} > 350~\mathrm{GeV}~~\mbox{at 95\% C.L.} \quad \mathrm{BR}(Q\to Wq)=100\%
\end{equation}

In \cite{Aad:2012bb} a search for pair production of $b^\prime$ decaying 100\% to $Wt$ is performed. Two of the 4 $W$s after top decay are then required to decay leptonically and the analysis considers final states containing two isolated same-sign leptons, $\geq$2 jets and large missing transverse momentum. The obtained bound is:
\begin{equation}
 m_{b^\prime} > 450~\mathrm{GeV}~~\mbox{at 95\% C.L.} \quad \mathrm{BR}(b^\prime\to Wt)=100\%
\end{equation}

In \cite{ATLAS:2012aw} a search for pair production of $b^\prime$ decaying 100\% to $Wt$ is performed. This search is very similar to \cite{Aad:2012bb}, but in this case only one of the 4 $W$s after top decay is then required to decay leptonically and the analysis considers final states containing exactly one lepton, $\geq$6 jets and large missing transverse momentum. The bound is however very close to the one obtained in \cite{Aad:2012bb}:
\begin{equation}
 m_{b^\prime} > 480~\mathrm{GeV}~~\mbox{at 95\% C.L.} \quad \mathrm{BR}(b^\prime\to Wt)=100\%
\end{equation}
This search has been reinterpreted in \cite{Rao:2012gf} to compute bounds on a VL top partner mixing only with the third SM generation: the obtained bounds are $m_T>415$ GeV or $m_T>557$ GeV for $BR(T\to Wb)=0\%$ or $BR(T\to Wb)=100\%$, respectively.

\subsubsection*{ATLAS @ $2~\mathrm{fb}^{-1}$}

In \cite{Aad:2012ak} a search for pair production of $b^\prime$ is performed. At least one $b^\prime$ is then required to decay into $Zb$, while the other can decay either to $Zb$ or $Wt$. A parameter which describes the fraction of events with at least one $b^\prime$ decaying to $Zb$ depending on the BR is defined: $\beta= 2 \times BR(b^\prime \to Zb) - BR(b^\prime \to Zb)^2$. The case $\beta=1$ corresponds to 100\% decay into $Zb$. The parameter $\beta$ is also computed for the case of a VL $b^\prime$ mixing to third generation in the mass range $m_{b^\prime}=200-700$ GeV, obtaining $\beta=0.9-0.5$. The final states considered in this search contain $\geq$2 opposite-sign leptons and one b-tagged jet. Limits are given both for 100\% decay assumption and for a specific VL scenario:
\begin{eqnarray}
 m_{b^\prime} > 400~\mathrm{GeV}~~\mbox{at 95\% C.L.} &\quad& \mathrm{BR}(b^\prime\to Zb)=100\% \\
 m_{b^\prime} > 358~\mathrm{GeV}~~\mbox{at 95\% C.L.} &\quad& \mbox{for VL singlet mixing only with third generation}
\end{eqnarray}

\subsubsection*{ATLAS @ $4.64~\mathrm{fb}^{-1}$}

In \cite{ATLAS-CONF-2012-137} a search for single production of VLQs coupling only to light generations is presented. The final state consists in a reconstructed vector boson (W or Z) decaying leptonically and 2 jets. The obtained bounds (which depend on the assumption on the VLQ couplings) are:
\begin{eqnarray}
 m_{b^\prime} > 1120~\mathrm{GeV}~~\mbox{at 95\% C.L.} &\quad& \mathrm{BR}(b^\prime\to Wq)=100\% \\
 m_{X_{5/3}} > 1420~\mathrm{GeV}~~\mbox{at 95\% C.L.} &\quad& \mathrm{BR}(X_{5/3}\to Wq)=100\% \\
 m_{t^\prime} > 1080~\mathrm{GeV}~~\mbox{at 95\% C.L.} &\quad& \mathrm{BR}(t^\prime\to Zt)=100\% 
\end{eqnarray}

\subsubsection*{ATLAS @ $4.7~\mathrm{fb}^{-1}$}

In \cite{Aad:2012uu}, a search for pair production of a heavy particle $T$ decaying to $tA_0$ where $A_0$ is an invisible (DM) particle is performed. The search is performed in the $t\bar t+ MET$ channel, where both the $W$s coming from tops decay leptonically. The final state considered in the analysis contain two leptons and missing transverse energy. Bounds are given for combinations of $T$ and $A_0$ masses, excluding heavy quarks up with massses $380\mbox{ GeV}<M_T<420$ GeV for stable neutral states with masses below 100 GeV.

In \cite{ATLAS:2012qe} a search for pair production of $t^\prime$ is performed. Depending on its branching ratios, the $t^\prime$ is assumed either to belong to a chiral fourth-generation or to be vector-like. The search looks for final states containing one isolated lepton with high $p_T$, at least 3 jets and large missing energy. Results are given either assuming $BR(t^\prime\to Wb)=100\%$, corresponding to the fourth-generation scenario, or considering $BR(t^\prime\to Wb)$ and $BR(t^\prime\to Ht)$ as independent parameters, with the relation $BR(t^\prime\to Zt)= 1-BR(t^\prime\to Wb)-BR(t^\prime\to Ht)$. The obtained bounds are:
\begin{eqnarray}
 m_{t^\prime} > 656~\mathrm{GeV}~~\mbox{at 95\% C.L.} &\quad& \mathrm{BR}(t^\prime\to Wb)=100\% \\
 m_{t^\prime} > 500~\mathrm{GeV}~~\mbox{at 95\% C.L.} &\quad& \mbox{for VL singlet with } \mathrm{BR}(t^\prime\to Wb)=50\% \mbox{ and } \mathrm{BR}(t^\prime\to Ht)=33\%
\end{eqnarray}
while the search is not sensitive to vector-like doublet scenarios.

In \cite{ATLAS-CONF-2012-130} a search in the same-sign dilepton channel is interpreted in terms of pair production of $b^\prime$ and single or pair production of the exotic $X_{5/3}$. Besides the 2 same sign leptons, the final state is required to contain at least 2 jets, with at least one b-jet, large missing energy and large $H_T$. The bounds considering only pair production processes are:
\begin{eqnarray}
 m_{b^\prime,X_{5/3}} > 670~\mathrm{GeV}~~\mbox{at 95\% C.L.} &\quad& \mathrm{BR}(b^\prime,X_{5/3}\to Wt)=100\%
\end{eqnarray}
while adding the single production of $X_{5/3}$ with different assumptions on its coupling with $Wt$, the bounds are:
\begin{eqnarray}
 m_{X_{5/3}} > 680-700~\mathrm{GeV}~~\mbox{at 95\% C.L.} &\quad& \mathrm{BR}(X_{5/3}\to Wt)=100\%
\end{eqnarray}

\subsubsection*{CMS @ $1.14~\mathrm{fb}^{-1}$}

In \cite{Chatrchyan:2011ay} a search for pair productoin of a VL $t^\prime$ quark is performed. The $t^\prime$ is assumed to decay 100\% into $Zt$ whereas the decay into $tH$ is supposed to be kinematically forbidden. The final state must contain $\geq$3 leptons and $\geq$2 jets. The obtained bound is:
\begin{equation}
 m_{t^\prime} > 475~\mathrm{GeV}~~\mbox{at 95\% C.L.} \quad \mathrm{BR}(t^\prime\to Zt)=100\%
\end{equation}

%

\subsubsection*{CMS @ $4.9~\mathrm{fb}^{-1}$}

In \cite{Chatrchyan:2012yea} a search for $b^\prime$ pair production is performed. The $b^\prime$ is then supposed to decay exclusively to $Wt$. The analysis searches for final states containing $\geq$2 electrons or at least one muon and $\geq$4 ($\geq$2) jets for same-sign dilepton (trilepton) events. The obtained bound is:
\begin{equation}
 m_{b^\prime} > 611~\mathrm{GeV}~~\mbox{at 95\% C.L.} \quad \mathrm{BR}(b^\prime\to Wt)=100\%
\end{equation}

In \cite{CMS-PAS-EXO-11-066} a search for $b^\prime$ pair production with subsequent $b^\prime\to Zb$ decay is performed. The decay branching ratio is assumed to be 100\% and the search focuses on final states with a pair of opposite charged leptons (to reconstruct the $Z$ boson) and at least one b-tagged jet. The obtained bound is:
\begin{equation}
 m_{b^\prime} > 550~\mathrm{GeV}~~\mbox{at 95\% C.L.} \quad \mathrm{BR}(b^\prime\to Zb)=100\%
\end{equation}

\subsubsection*{CMS @ $5~\mathrm{fb}^{-1}$}

In \cite{CMS:2012ab} a search for pair production of $t^\prime$ is performed. The $t^\prime$ decays exclusively to $Wb$ and the $W$ bosons are both required to decay leptonically. Events are required to have two opposite-sign leptons, $\geq$2 jets, of which exactly two coming from bottom fragmentation. The obtained bound is:
\begin{equation}
 m_{t^\prime} > 557~\mathrm{GeV}~~\mbox{at 95\% C.L.} \quad \mathrm{BR}(t^\prime\to Wb)=100\%
\end{equation}

In \cite{CMS-PAS-B2G-12-003} a search for pair production of exotic $X_{5/3}$ is performed. The analysis considers final states composed of exactly 2 same sign leptons, at least 4 jets and large $H_T$. The obtained bound is:
\begin{equation}
 m_{X_{5/3}} > 645~\mathrm{GeV}~~\mbox{at 95\% C.L.} \quad \mathrm{BR}(X_{5/3}\to Wt)=100\%
\end{equation}

Two searches have been performed on the same final state, containing a single isolated lepton ($e$ or $\mu$), large missing transverse energy, $\geq$4 jets, one of which originating from the fragmentation of a bottom quark, though considering different kinematical cuts. In \cite{Chatrchyan:2012vu} a search for $t^\prime$ pair production is performed, where the $t^\prime$ is supposed to decay 100\% to $Wb$, while in \cite{Chatrchyan:2012af} a search of pair production of heavy quarks decaying either to $Zt$ or $Wt$ is presented. The combined bounds is:
\begin{eqnarray}
 m_{t^\prime} > 560~\mathrm{GeV}~~\mbox{at 95\% C.L.} &\quad& \mathrm{BR}(t^\prime\to Wb)=100\%\\
 m_{t^\prime} > 625~\mathrm{GeV}~~\mbox{at 95\% C.L.} &\quad& \mathrm{BR}(t^\prime\to Zt)=100\%\\
 m_{b^\prime} > 675~\mathrm{GeV}~~\mbox{at 95\% C.L.} &\quad& \mathrm{BR}(b^\prime\to Wt)=100\%
\end{eqnarray}

\section{Signatures of vector-like quarks at LHC}
\label{Sect:signatures}

\subsection{A general picture}
The identification of the channels which may lead to the discovery of VLQs at the LHC depends on the scenario under consideration. In general, processes dominated by QCD, such as pair production, have the advantage of being model independent, while single production is driven by model-dependent processes. However, pair production suffers from a larger phase-space suppression with respect to single production, and if the VLQ mass is large enough, single production dominates over pair production. The VLQ mass corresponding to the equivalence between pair and single production cross sections depends on the specific model.
Excluding purely QCD processes, the production of VLQs is related to the interaction of the new states with SM particles. If VLQs interact with SM quarks through Yukawa couplings, a mixing is induced between quarks of different families, giving rise to FCNCs. 
On the other hand, in scenarios such as minimal universal extra-dimensions, the KK-odd VLQs do not mix with SM quarks and therefore they can only be produced in pairs or together with another KK-odd state.

\begin{figure}[ht]
\begin{center}
 \begin{picture}(75,40)(0,-10)
  \SetWidth{1}
  \Line[arrow,arrowpos=0.5,arrowlength=2,arrowwidth=1,arrowinset=0.2](10,20)(20,10)
  \Text(8,20)[rc]{\small{$q$}}
  \Line[arrow,arrowpos=0.5,arrowlength=2,arrowwidth=1,arrowinset=0.2](20,10)(10,0)
  \Text(8,0)[rc]{\small{$\bar q$}}
  \Gluon(20,10)(40,10){2}{4}
  \Text(30,20)[tc]{\small{$g$}}
  \Line[arrow,arrowpos=0.5,arrowlength=2,arrowwidth=1,arrowinset=0.2](40,10)(50,20)
  \Text(52,20)[lc]{\small{$Q_{VL}$}}
  \Line[arrow,arrowpos=0.5,arrowlength=2,arrowwidth=1,arrowinset=0.2](50,0)(40,10)
  \Text(52,0)[lc]{\small{$\bar Q_{VL}$}}
 \end{picture}
 \begin{picture}(75,40)(0,-10)
  \SetWidth{1}
  \Gluon(10,20)(20,10){2}{3}
  \Text(8,20)[rc]{\small{$g$}}
  \Gluon(10,0)(20,10){2}{3}
  \Text(8,0)[rc]{\small{$g$}}
  \Gluon(20,10)(40,10){2}{4}
  \Text(30,20)[tc]{\small{$g$}}
  \Line[arrow,arrowpos=0.5,arrowlength=2,arrowwidth=1,arrowinset=0.2](40,10)(50,20)
  \Text(52,20)[lc]{\small{$Q_{VL}$}}
  \Line[arrow,arrowpos=0.5,arrowlength=2,arrowwidth=1,arrowinset=0.2](50,0)(40,10)
  \Text(52,0)[lc]{\small{$\bar Q_{VL}$}}
 \end{picture}
 \begin{picture}(75,40)(0,-10)
  \SetWidth{1}
  \Gluon(10,20)(30,20){2}{4}
  \Text(8,20)[rc]{\small{$g$}}
  \Gluon(10,0)(30,00){2}{4}
  \Text(8,0)[rc]{\small{$g$}}
  \Line[arrow,arrowpos=0.5,arrowlength=2,arrowwidth=1,arrowinset=0.2](30,0)(30,20)
  \Text(34,10)[cl]{\small{$Q_{VL}$}}
  \Line[arrow,arrowpos=0.5,arrowlength=2,arrowwidth=1,arrowinset=0.2](30,20)(50,20)
  \Text(52,20)[lc]{\small{$Q_{VL}$}}
  \Line[arrow,arrowpos=0.5,arrowlength=2,arrowwidth=1,arrowinset=0.2](50,0)(30,0)
  \Text(52,0)[lc]{\small{$\bar Q_{VL}$}}
 \end{picture}
\end{center}

\begin{center}
 \begin{picture}(75,40)(0,-10)
  \SetWidth{1}
  \Line[arrow,arrowpos=0.5,arrowlength=2,arrowwidth=1,arrowinset=0.2](10,20)(20,10)
  \Text(8,20)[rc]{\small{$q_i$}}
  \Line[arrow,arrowpos=0.5,arrowlength=2,arrowwidth=1,arrowinset=0.2](20,10)(10,0)
  \Text(8,0)[rc]{\small{$\bar q_j$}}
  \Photon(20,10)(40,10){2}{4}
  \Text(30,20)[tc]{\small{$V$}}
  \Line[arrow,arrowpos=0.5,arrowlength=2,arrowwidth=1,arrowinset=0.2](40,10)(50,20)
  \Text(52,20)[lc]{\small{$Q_{VL}$}}
  \Line[arrow,arrowpos=0.5,arrowlength=2,arrowwidth=1,arrowinset=0.2](50,0)(40,10)
  \Text(52,0)[lc]{\small{$\bar Q_{VL}$}}
 \end{picture}
 \begin{picture}(75,40)(0,-10)
  \SetWidth{1}
  \Line[](10,20)(30,20)
  \Text(8,20)[rc]{\small{$q_i$}}
  \Line[](30,0)(10,0)
  \Text(8,0)[rc]{\small{$q_j$}}
  \Photon(30,20)(30,0){2}{4}
  \Text(34,10)[cl]{\small{$V$}}
  \Line[](30,20)(50,20)
  \Text(52,20)[lc]{\small{$Q_{VL}$}}
  \Line[](50,0)(30,0)
  \Text(52,0)[lc]{\small{$Q_{VL}$}}
 \end{picture}
 \begin{picture}(75,40)(0,-10)
  \SetWidth{1}
  \Line[arrow,arrowpos=0.5,arrowlength=2,arrowwidth=1,arrowinset=0.2](10,20)(20,10)
  \Text(8,20)[rc]{\small{$q_i$}}
  \Line[arrow,arrowpos=0.5,arrowlength=2,arrowwidth=1,arrowinset=0.2](20,10)(10,0)
  \Text(8,0)[rc]{\small{$\bar q_j$}}
  \Line[dash,dashsize=2.5](20,10)(40,10)
  \Text(30,20)[tc]{\small{$S$}}
  \Line[arrow,arrowpos=0.5,arrowlength=2,arrowwidth=1,arrowinset=0.2](40,10)(50,20)
  \Text(52,20)[lc]{\small{$Q_{VL}$}}
  \Line[arrow,arrowpos=0.5,arrowlength=2,arrowwidth=1,arrowinset=0.2](50,0)(40,10)
  \Text(52,0)[lc]{\small{$\bar Q_{VL}$}}
 \end{picture}
 \begin{picture}(75,40)(0,-10)
  \SetWidth{1}
  \Line[](10,20)(30,20)
  \Text(8,20)[rc]{\small{$q_i$}}
  \Line[](30,0)(10,0)
  \Text(8,0)[rc]{\small{$q_j$}}
  \Line[dash,dashsize=2.5](30,20)(30,0)
  \Text(34,10)[cl]{\small{$S$}}
  \Line[](30,20)(50,20)
  \Text(52,20)[lc]{\small{$Q_{VL}$}}
  \Line[](50,0)(30,0)
  \Text(52,0)[lc]{\small{$Q_{VL}$}}
 \end{picture}
\end{center}
\caption{Feynman diagrams for pair production of a generic VLQ. Above the dominant and model-independent QCD contributions, below the subdominant and model-dependent EW contributions. Arrows on fermion lines have been removed to account for both particles and antiparticles, when necessary. Notice the possibility to have FCNCs between SM quarks in the $V$ and $S$ s-channel diagram, which is peculiar to VL scenarios.}
\label{Fig.PairVL}
\end{figure}
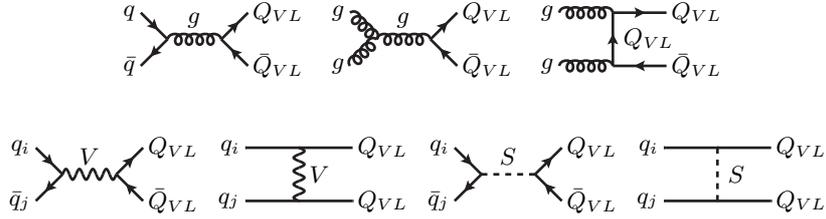

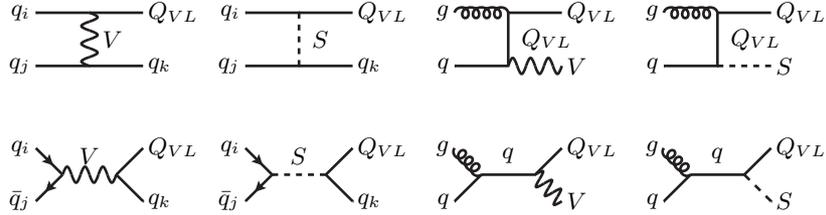
\begin{figure}[ht]
\begin{center}
 \begin{picture}(75,40)(0,-10)
  \SetWidth{1}
  \Line[](10,20)(30,20)
  \Text(8,20)[rc]{\small{$q_i$}}
  \Line[](10,0)(30,0)
  \Text(8,0)[rc]{\small{$q_j$}}
  \Photon(30,20)(30,0){3}{3}
  \Text(35,10)[lc]{\small{$V$}}
  \Line[](30,20)(50,20)
  \Text(52,20)[lc]{\small{$Q_{VL}$}}
  \Line[](30,0)(50,0)
  \Text(52,0)[lc]{\small{$q_k$}}
 \end{picture}
 \begin{picture}(75,40)(0,-10)
  \SetWidth{1}
  \Line[](10,20)(30,20)
  \Text(8,20)[rc]{\small{$q_i$}}
  \Line[](10,0)(30,0)
  \Text(8,0)[rc]{\small{$q_j$}}
  \Line[dash,dashsize=2.5](30,20)(30,0)
  \Text(35,10)[lc]{\small{$S$}}
  \Line[](30,20)(50,20)
  \Text(52,20)[lc]{\small{$Q_{VL}$}}
  \Line[](30,0)(50,0)
  \Text(52,0)[lc]{\small{$q_k$}}
 \end{picture}
 \begin{picture}(75,40)(0,-10)
  \SetWidth{1}
  \Line[](10,0)(30,0)
  \Text(8,0)[rc]{\small{$q$}}
  \Gluon(10,20)(30,20){2}{4}
  \Text(8,20)[rc]{\small{$g$}}
  \Line[](30,0)(30,20)
  \Text(35,10)[lc]{\small{$Q_{VL}$}}
  \Photon(30,0)(50,0){3}{3}
  \Text(52,0)[lc]{\small{$V$}}
  \Line[](30,20)(50,20)
  \Text(52,20)[lc]{\small{$Q_{VL}$}}
 \end{picture}
 \begin{picture}(75,40)(0,-10)
  \SetWidth{1}
  \Line[](10,0)(30,0)
  \Text(8,0)[rc]{\small{$q$}}
  \Gluon(10,20)(30,20){2}{4}
  \Text(8,20)[rc]{\small{$g$}}
  \Line[](30,0)(30,20)
  \Text(35,10)[lc]{\small{$Q_{VL}$}}
  \Line[dash,dashsize=2.5](30,0)(50,0)
  \Text(52,0)[lc]{\small{$S$}}
  \Line[](30,20)(50,20)
  \Text(52,20)[lc]{\small{$Q_{VL}$}}
 \end{picture}
\end{center}

\begin{center}
 \begin{picture}(75,40)(0,-10)
  \SetWidth{1}
  \Line[arrow,arrowpos=0.5,arrowlength=2,arrowwidth=1,arrowinset=0.2](10,20)(20,10)
  \Text(8,20)[rc]{\small{$q_i$}}
  \Line[arrow,arrowpos=0.5,arrowlength=2,arrowwidth=1,arrowinset=0.2](20,10)(10,0)
  \Text(8,0)[rc]{\small{$\bar q_j$}}
  \Photon(20,10)(40,10){3}{3}
  \Text(30,20)[tc]{\small{$V$}}
  \Line[](40,10)(50,20)
  \Text(52,20)[lc]{\small{$Q_{VL}$}}
  \Line[](50,0)(40,10)
  \Text(52,0)[lc]{\small{$q_k$}}
 \end{picture}
 \begin{picture}(75,40)(0,-10)
  \SetWidth{1}
  \Line[arrow,arrowpos=0.5,arrowlength=2,arrowwidth=1,arrowinset=0.2](10,20)(20,10)
  \Text(8,20)[rc]{\small{$q_i$}}
  \Line[arrow,arrowpos=0.5,arrowlength=2,arrowwidth=1,arrowinset=0.2](20,10)(10,0)
  \Text(8,0)[rc]{\small{$\bar q_j$}}
  \Line[dash,dashsize=2.5](20,10)(40,10)
  \Text(30,20)[tc]{\small{$S$}}
  \Line[](40,10)(50,20)
  \Text(52,20)[lc]{\small{$Q_{VL}$}}
  \Line[](50,0)(40,10)
  \Text(52,0)[lc]{\small{$q_k$}}
 \end{picture}
 \begin{picture}(75,40)(0,-10)
  \SetWidth{1}
  \Gluon(20,10)(10,20){2}{3}
  \Text(8,20)[rc]{\small{$g$}}
  \Line[](10,0)(20,10)
  \Text(8,0)[rc]{\small{$q$}}
  \Line[](20,10)(40,10)
  \Text(30,20)[tc]{\small{$q$}}
  \Line[](40,10)(50,20)
  \Text(52,20)[lc]{\small{$Q_{VL}$}}
  \Photon(50,0)(40,10){3}{3}
  \Text(52,0)[lc]{\small{$V$}}
 \end{picture}
 \begin{picture}(75,40)(0,-10)
  \SetWidth{1}
  \Gluon(20,10)(10,20){2}{3}
  \Text(8,20)[rc]{\small{$g$}}
  \Line[](10,0)(20,10)
  \Text(8,0)[rc]{\small{$q$}}
  \Line[](20,10)(40,10)
  \Text(30,20)[tc]{\small{$q$}}
  \Line[](40,10)(50,20)
  \Text(52,20)[lc]{\small{$Q_{VL}$}}
  \Line[dash,dashsize=2.5](50,0)(40,10)
  \Text(52,0)[lc]{\small{$S$}}
 \end{picture}
\end{center}
\caption{Feynman diagrams for single production of a generic VLQ. VLQs can interact with SM quarks both through charged currents and neutral currents, allowing FCNCs also within SM states in diagrams with $q_i-q_j-\{V,S\}$ interactions. Arrows on fermion lines have been removed to account for both particles and antiparticles, when necessary. Notice that not all diagrams are allowed for a specific VL quark (e.g. neutral currents are not allowed for quarks with exotic electric charges).}
\label{Fig.SingleVL}
\end{figure}

The Feynman diagrams for pair and single production of VLQs are shown in Figs.~\ref{Fig.PairVL}-~\ref{Fig.SingleVL}. 

The decay channels of VLQs are model-dependent too, and this is the most relevant problem when trying to interpret experimental bounds on new heavy quarks, due to the fact that these bounds are generally obtained under strong assumptions on the branching ratios of the new states. 
In the following, the decay channels for each VLQ will be analysed to find which final states are possible if VLQs are produced at the LHC. Of course this analysis cannot be completely general, because VLQ couplings are model-dependent. VLQs are therefore assumed to interact with SM quarks of all flavours through Yukawa couplings. This is the minimal scenario of new physics with VLQs, in which they are the only new states besides SM particles. Next-to-minimal scenarios with the addition of more than one VLQ representation or in which VLQs interact with SM particles and another invisible particle (as in UED scenarios) will not be considered. 

\begin{figure}
\centering
\epsfig{file=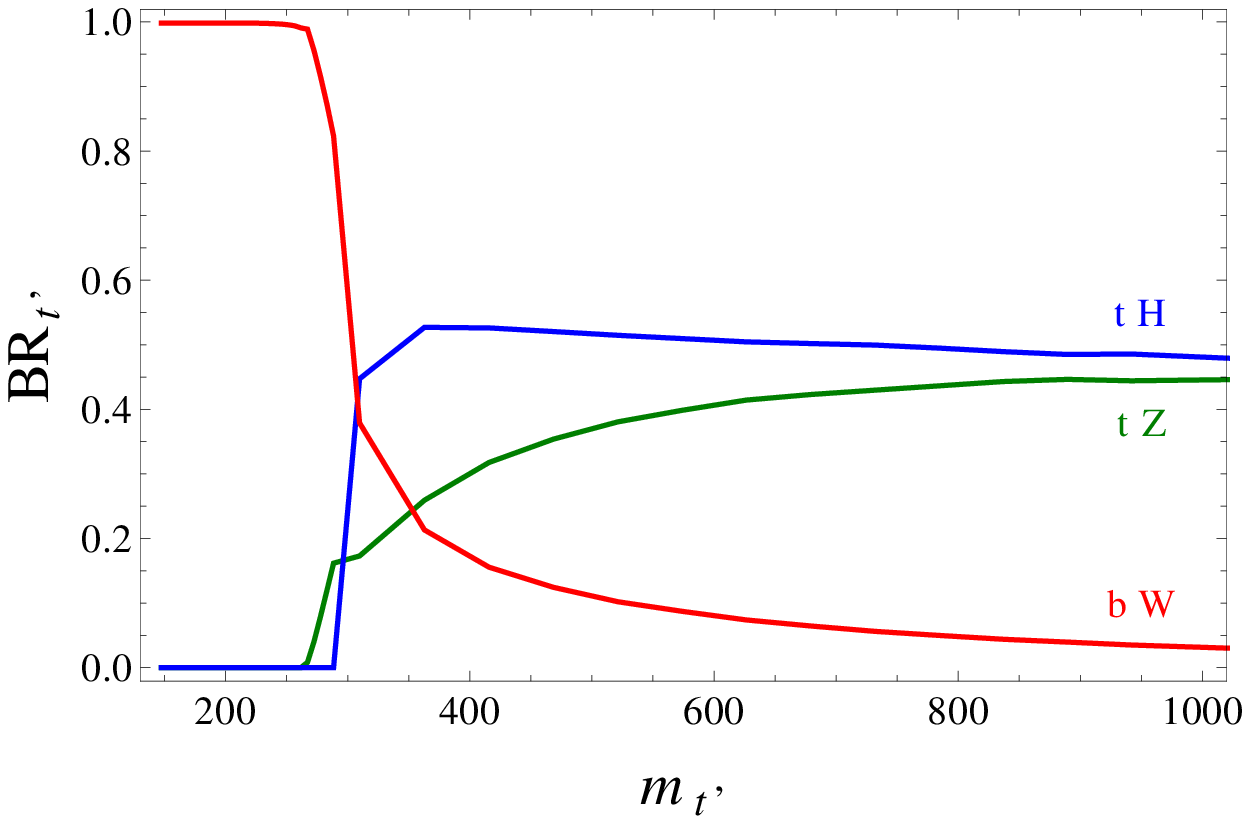,width=.48\textwidth}\hfill
\epsfig{file=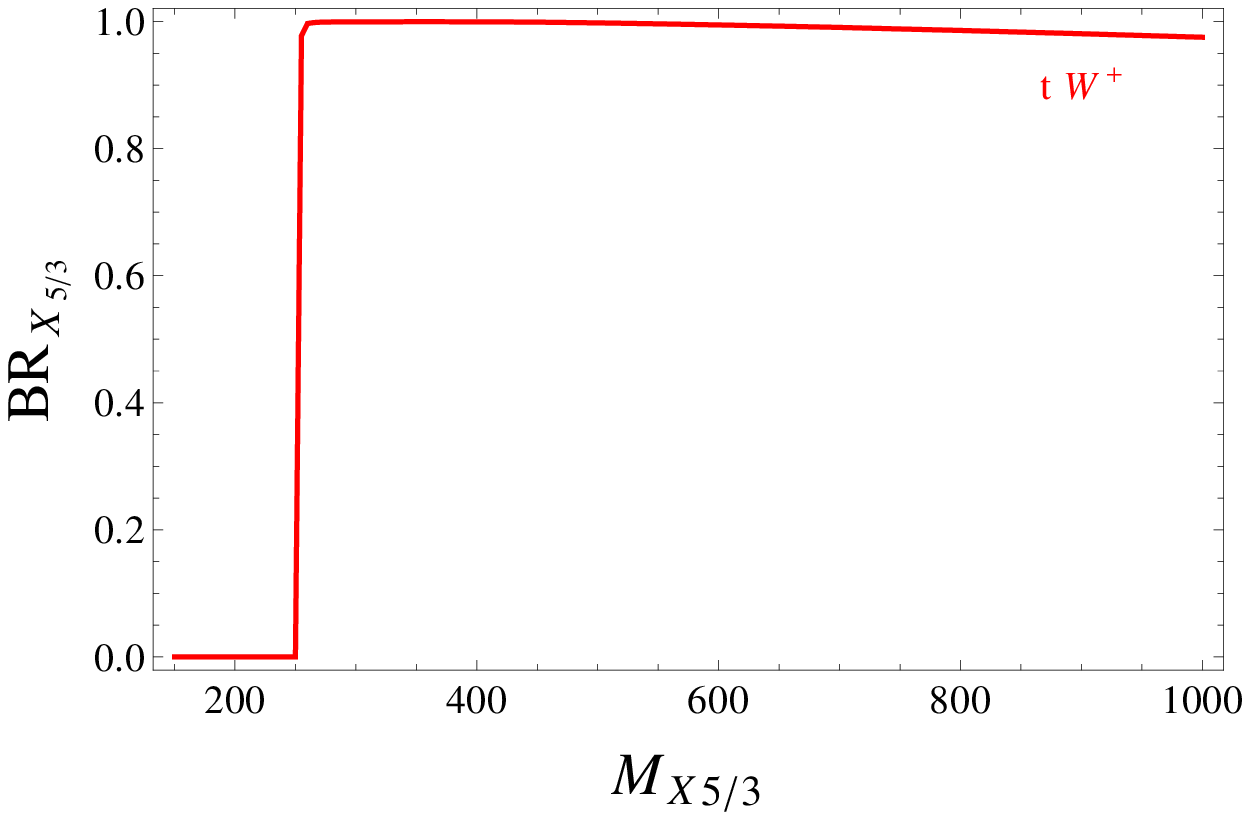,width=.48\textwidth}\\[5pt]
\epsfig{file=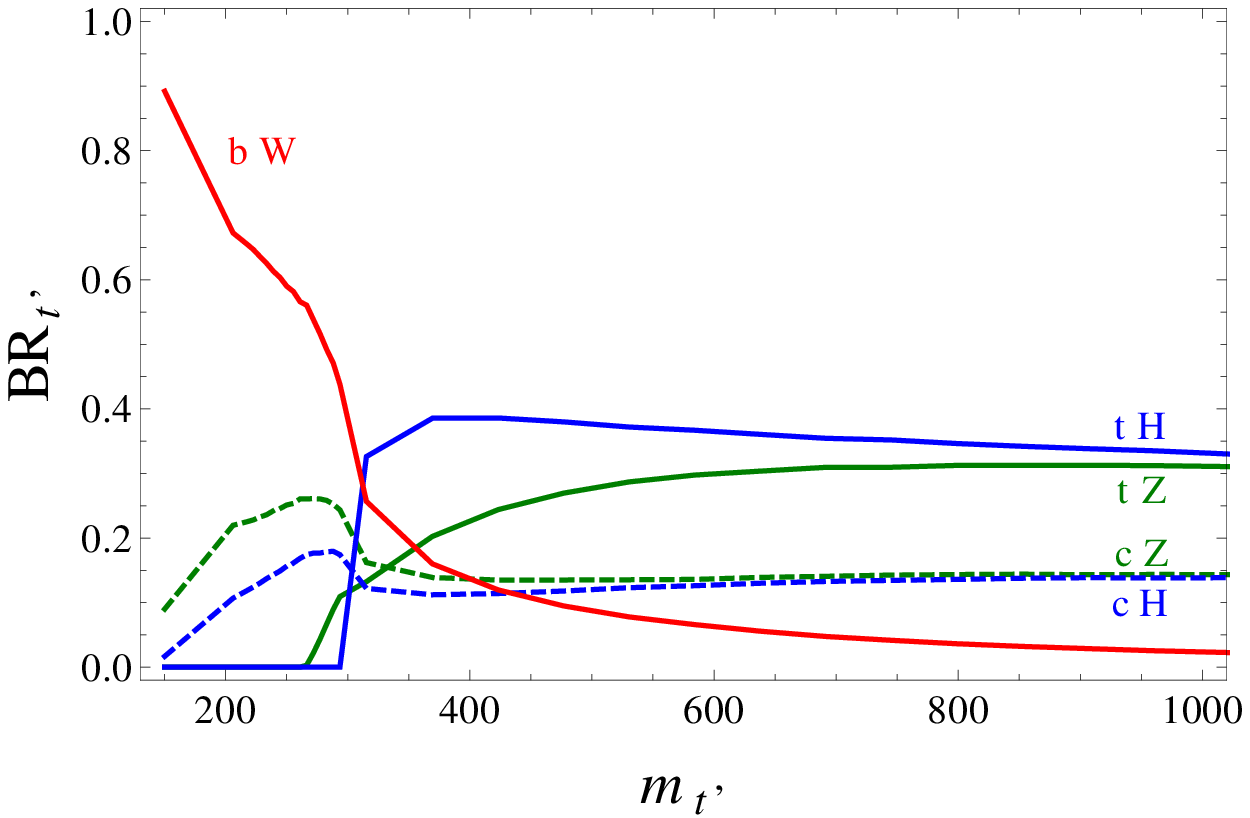,width=.48\textwidth}\hfill
\epsfig{file=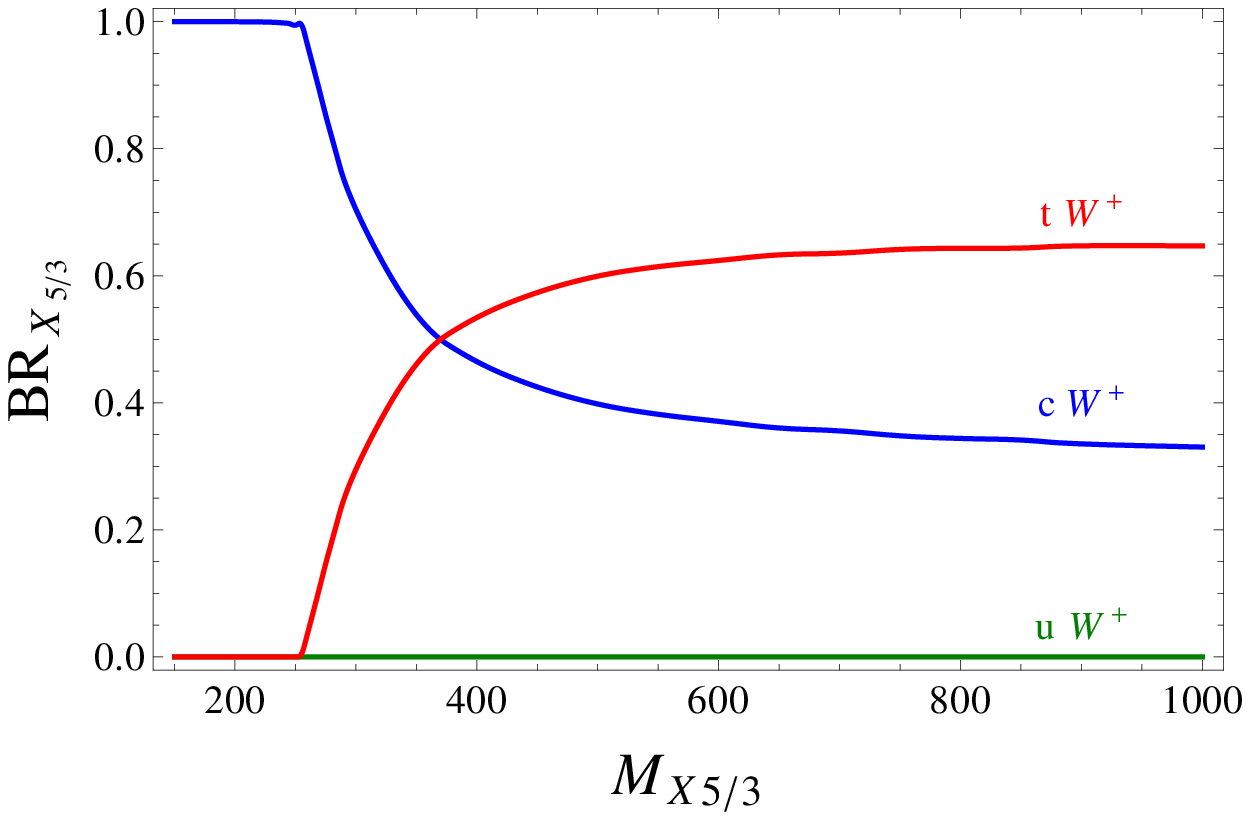,width=.48\textwidth}\\[5pt]
\epsfig{file=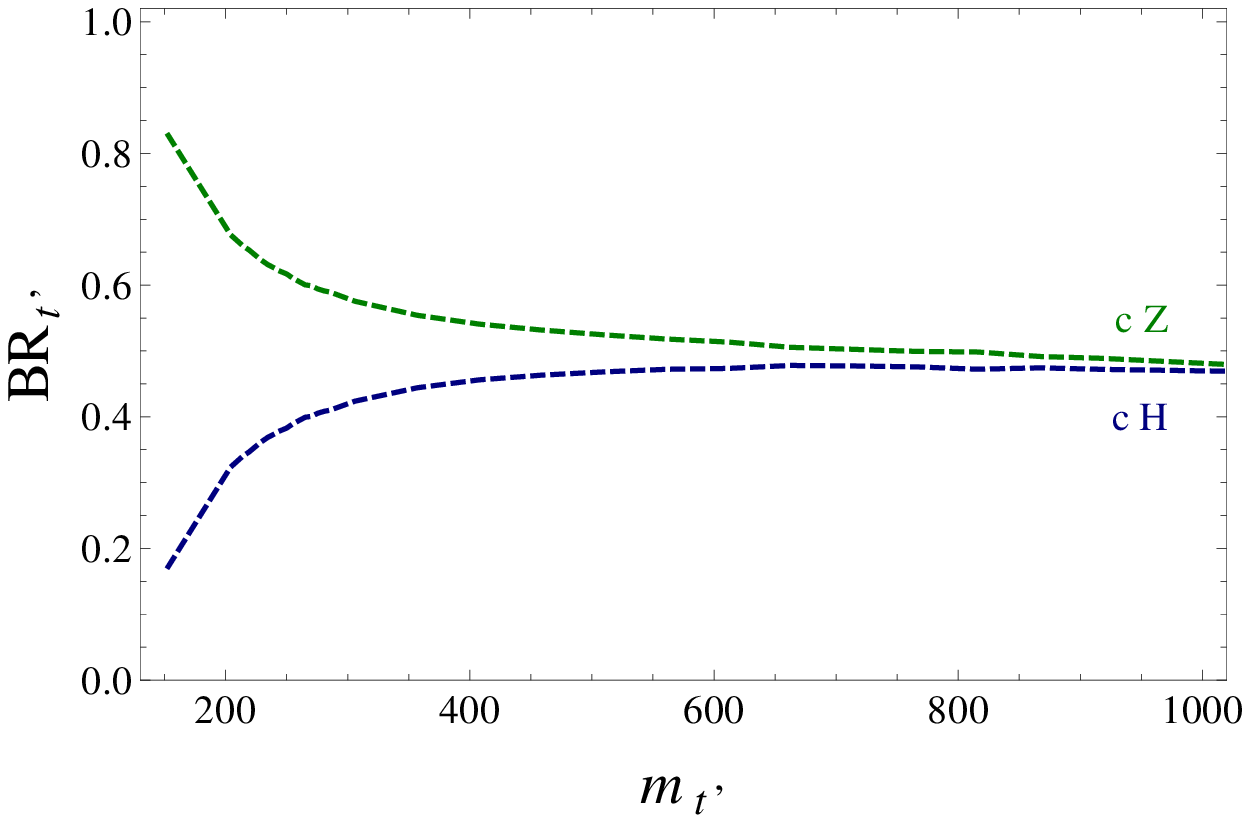,width=.48\textwidth}\hfill
\epsfig{file=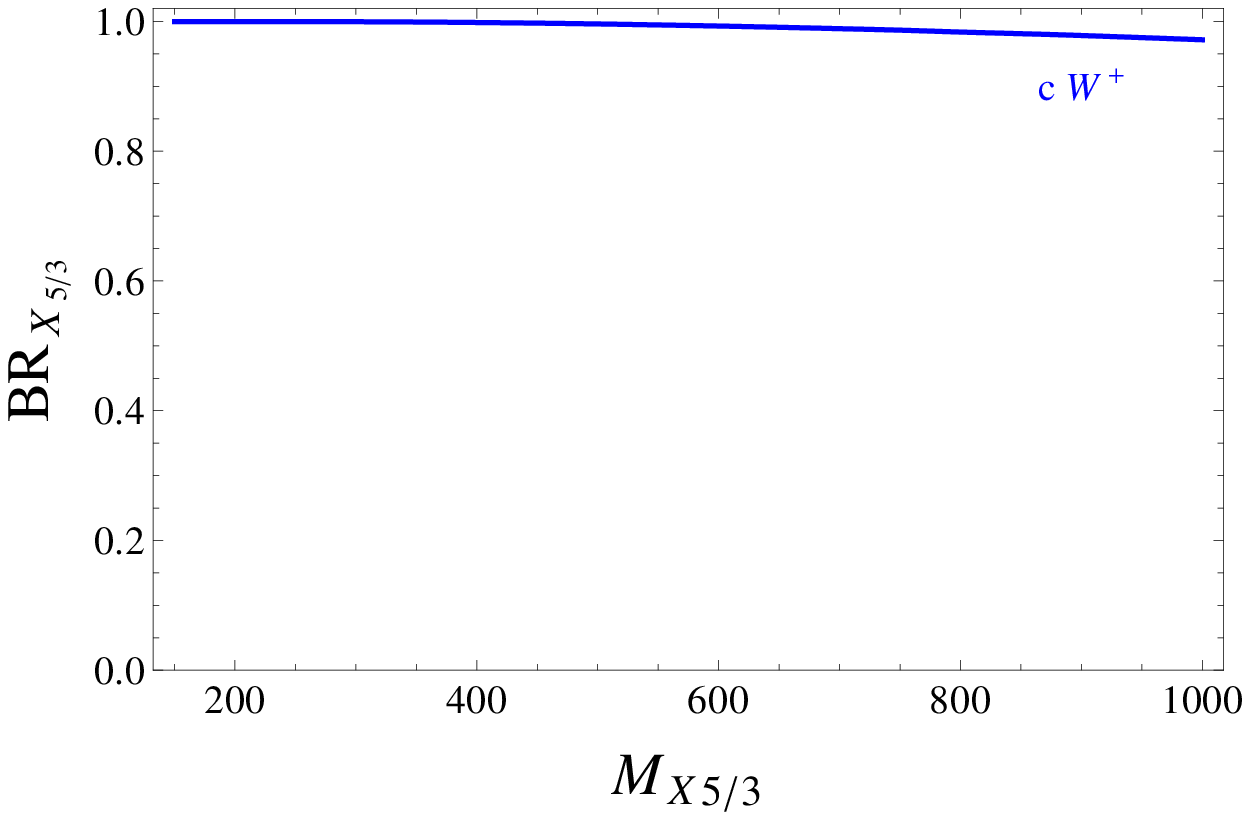,width=.48\textwidth}
\caption{Branching ratios for $t^\prime$ (first column) and $X_{5/3}$ (second column) belonging to the non-SM doublet $(X_{5/3},t^\prime)$ for different mixing hypotheses. From top to bottom: mixing only with top; mixing with top and charm; mixing only with charm. The mixing parameters have been saturated according to the bounds obtained in \cite{Cacciapaglia:2011fx}. Notice that the BR of $X_{5/3}$ in the case of mixing only with top or charm is not identically 1: this is due to the presence of three-body decays, which have not been included in the plot.}
\label{BRstprimeX}
\end{figure}

\subsubsection{$t^\prime$ and $b^\prime$}

\begin{table}
\centering
\begin{tabular}{c|c|c|c}
\multirow{2}{*}{Charge} & \multirow{2}{*}{Resonant state} & \multicolumn{1}{c}{After $t^\prime$ decay} & \multicolumn{1}{c}{After $b^\prime$ decay} \\
 & & \multicolumn{2}{c}{(including decays to other VLQs, if allowed)} \\
\toprule
0 & \multirow{4}{*}{$\begin{array}{c}t^\prime \bar t^\prime \\ b^\prime \bar b^\prime\end{array}$} 
  & $u_i \bar u_j + \{ZZ,ZH,HH\}$                         & $d_i \bar d_j + \{ZZ,ZH,HH\}$ \\
& & $\{d_i \bar d_j,X_{5/3} \bar X_{5/3}\} + W^+W^-$      & $\{u_i \bar u_j,Y_{-4/3} \bar Y_{-4/3}\} + W^+W^-$      \\
& & $u_i \bar d_j + W^- + \{Z,H\} \mbox{ (and c.c.)}$     & $d_i \bar u_j + W^+ + \{Z,H\} \mbox{ (and c.c.)}$     \\
& & $u_i \bar X_{5/3} + W^+ + \{Z,H\} \mbox{ (and c.c.)}$ & $d_i \bar Y_{-4/3} + W^- + \{Z,H\} \mbox{ (and c.c.)}$ \\
\cmidrule(r){2-4}
  & \multirow{3}{*}{$\begin{array}{c} t^\prime \bar u_i \\ b^\prime \bar d_i \end{array}$} 
  & $u_j \bar u_i + \{Z,H\}$ & $d_j \bar d_i + \{Z,H\}$ \\
& & $d_j \bar u_i + W^+$     & $u_j \bar d_i + W^-$     \\
& & $X_{5/3} \bar u_i + W^-$ & $Y_{-4/3} \bar d_i + W^+$ \\
\midrule
1/3 & \multirow{3}{*}{$\begin{array}{c} t^\prime d_i \\ b^\prime u_i \end{array}$} 
    & $u_j d_i + \{Z,H\}$ & $d_j u_i + \{Z,H\}$ \\
&   & $d_j d_i + W^+$     & $u_j u_i + W^-$     \\
&   & $X_{5/3} d_i + W^-$ & $Y_{-4/3} u_i + W^+$ \\
\cmidrule(r){2-4}
    & \multirow{4}{*}{$t^\prime b^\prime$} 
    & \multicolumn{2}{c}{$u_i d_j + \{ZZ,ZH,HH\}$} \\
&   & \multicolumn{2}{c}{$u_i u_j + W^- +\{Z,H\} \quad d_i d_j + W^+ +\{Z,H\}$} \\
&   & \multicolumn{2}{c}{$u_i Y_{-4/3} + W^+ +\{Z,H\} \quad X_{5/3} d_i + W^- +\{Z,H\}$} \\
&   & \multicolumn{2}{c}{$d_i u_j + W^+W^-$} \\
\cmidrule(r){2-4}
    & \multirow{3}{*}{$\begin{array}{c} \bar t^\prime W^+ \\ \bar b^\prime Z \quad \bar b^\prime H \end{array}$}
    & $\bar u_i + W^- + \{Z,H\}$ & $\bar d_i + \{ZZ,ZH,HH\}$      \\
&   & $\bar d_i + W^+W^- $       & $\bar u_i + W^+ + \{Z,H\}$     \\
&   & $\bar X_{5/3} + W^-W^- $   & $\bar Y_{-4/3} + W^- + \{Z,H\}$\\
\midrule
2/3 & \multirow{3}{*}{$\bar b^\prime \bar b^\prime$}
&   & $\bar d_i \bar d_j + \{ZZ,ZH,HH\}$      \\
& & & $\bar u_i \bar u_j + W^+W^+$            \\
& & & $\bar Y_{-4/3} \bar Y_{-4/3} + W^-W^-$     \\
& & & $\bar d_i \bar u_j + W^+ + \{Z,H\}$     \\
& & & $\bar d_i \bar Y_{-4/3} + W^- + \{Z,H\}$ \\
\cmidrule(r){2-4}
    & \multirow{3}{*}{$\bar b^\prime \bar d_i$}  
&   & $\bar d_j \bar d_i + \{Z,H\}$ \\
& & & $\bar u_j \bar d_i + W^+$     \\
& & & $\bar Y_{-4/3} \bar d_i + W^-$ \\
\cmidrule(r){2-4}
    & \multirow{3}{*}{$\begin{array}{c} t^\prime Z \quad t^\prime H \\ b^\prime W^+ \end{array}$}
    & $u_i + \{ZZ,ZH,HH\}$      & $d_i + W^+ + \{Z,H\}$ \\
&   & $d_i + W^+ + \{Z,H\}$     & $u_i + W^+W^- $       \\
&   & $X_{5/3} + W^- + \{Z,H\}$ & $Y_{-4/3} + W^+W^+ $  \\
\midrule
1 & \multirow{3}{*}{$\begin{array}{c} t^\prime \bar d_i \\ \bar b^\prime u_i \end{array}$} 
  & $u_j \bar d_i + \{Z,H\}$ & $\bar d_j u_i + \{Z,H\}$ \\
& & $d_j \bar d_i + W^+$     & $\bar u_j u_i + W^+$     \\
& & $X_{5/3} \bar d_i + W^-$ & $\bar Y_{-4/3} u_i + W^-$ \\
\cmidrule(r){2-4}
& \multirow{4}{*}{$t^\prime \bar b^\prime$} 
    & \multicolumn{2}{c}{$u_i \bar d_j + \{ZZ,ZH,HH\}$} \\
&   & \multicolumn{2}{c}{$u_i \bar u_j + W^+ +\{Z,H\} \quad d_i \bar d_j + W^+ +\{Z,H\}$} \\
&   & \multicolumn{2}{c}{$u_i \bar Y_{-4/3} + W^- +\{Z,H\} \quad X_{5/3} \bar d_i + W^- +\{Z,H\}$} \\
&   & \multicolumn{2}{c}{$d_i \bar u_j + W^+W^+$} \\
\midrule
4/3 & \multirow{5}{*}{$t^\prime t^\prime$} 
    & $u_i u_j + \{ZZ,ZH,HH\}$      \\
&   & $d_i d_j + W^+W^+$            \\
&   & $X_{5/3}X_{5/3} + W^-W^-$     \\
&   & $u_i d_j + W^+ + \{Z,H\}$     \\
&   & $u_i X_{5/3} + W^- + \{Z,H\}$ \\
\cmidrule(r){2-4}
    & \multirow{3}{*}{$t^\prime u_i$}  
    & $u_j u_i + \{Z,H\}$ \\
&   & $d_j u_i + W^+$     \\
&   & $X_{5/3} u_i + W^-$ \\
\cmidrule(r){2-4}
    & \multirow{3}{*}{$\bar b^\prime W^+$}  
  & & $\bar d_i + W^+ + \{Z,H\}$ \\
& & & $\bar u_i + W^+W^+$     \\
& & & $\bar Y_{-4/3} + W^+W^-$ \\
\toprule
\end{tabular}
\caption{Production and decay channels for $t^\prime$ and $b^\prime$. Only configurations with $Q\geq0$ have been shown for simplicity. Throughout the table, $u_i=u,c,t(,t^\prime)$ and $d_i=d,s,b(,b^\prime)$.}
\label{tbprimeproduction}
\end{table}

Being partners of SM quarks, the $\{t^\prime,b^\prime\}$ can be produced exactly in the same way as any SM quark, but for the presence of FCNC channels. On the other hand, depending on their interactions with other SM quarks and on the representation they belongs to, the $\{t^\prime,b^\prime\}$ have different decay channels:
\begin{eqnarray}
 t^\prime &\to& \left\{\begin{array}{llr} d_i W^+,~X_{5/3} W^- & \mbox{with } d_i=d,s,b,b^\prime & CC \\ u_i Z,~u_i H & \mbox{with } u_i=u,c,t & NC \end{array}\right.\\
 b^\prime &\to& \left\{\begin{array}{llr} u_i W^-,~Y_{-4/3} W^+ & \mbox{with } u_i=u,c,t,t^\prime & CC \\ d_i Z,~d_i H & \mbox{with } d_i=d,s,b & NC \end{array}\right.
\end{eqnarray}
Of course not all decays may be kinematically allowed for a given scenario, but will depend on masses and couplings of the VLQs under consideration.
A key point which has not always been considered in experimental searches of top or bottom partners, is the possibility for VLQs to mix with all SM families: VLQs can decay in different channels, with branching ratios which depend on the relative strenght of the mixing between VLQs and SM quarks. The variation of the BRs of $t^\prime$ decay into $Vq$ is shown in Fig.~\ref{BRstprimeX}, which refers to a scenario in which the $t^\prime$ belongs to a non-SM doublet $(X_{5/3},t^\prime)$, analysed in detail in \cite{Cacciapaglia:2011fx}. 
The assumption that the $t^\prime$ mixes only with the top quark is therefore very strong from an experimental point of view: cuts optimized for scenarios with $BR(t^\prime\to Zt)=100\%$ or $BR(t^\prime\to Wb)=100\%$ may completely reject events coming from different channels, thus overestimating the bounds.

Limiting ourselves to the production of two particle states, it is possible to identify all the allowed channels for $\{t^\prime,b^\prime\}$ production and decay. The full list is shown in Tab.~\ref{tbprimeproduction}, where the channels are ordered in terms of the possible charge combinations allowed by a proton-proton interaction.

Even if the list is limited to two particle intermediate states, it is still possible to notice that if the $\{t^\prime,b^\prime\}$ is heavy enough to decay into a top quark or to other VLQs, there are many combinations which can result in multileptonic final states, allowing an easy identification of a $\{t^\prime,b^\prime\}$ signal against the SM background.
The multileptonic final states can be also enhanced by the presence of neutral currents, which are forbidden for a sequential chiral fourth family.

\subsubsection{$X_{5/3}$ and $Y_{-4/3}$}

\begin{table}
\centering
\begin{tabular}{c|c|c|c}
\multirow{2}{*}{Charge} & \multirow{2}{*}{Resonant state} & \multicolumn{1}{c}{After $X_{5/3}$ decay} & \multicolumn{1}{c}{After $Y_{-4/3}$ decay} \\
 & & \multicolumn{2}{c}{(including decays to other VLQs, if allowed)} \\
\toprule
0 & $\begin{array}{c}X_{5/3} \bar X_{5/3} \\ Y_{-4/3} \bar Y_{-4/3} \end{array}$ & $u_i \bar u_i + W^+W^-$ & $d_i \bar d_i + W^+W^-$ \\
\midrule
1/3 & $\bar Y_{-4/3} W^-$ & & $\bar d_i + W^+W^-$ \\
\midrule
2/3 & $\begin{array}{c}X_{5/3} W^- \\ \bar Y_{-4/3} \bar u_i \end{array}$ & $u_i + W^+W^-$ & $\bar d_i \bar u_i + W^+$ \\
\midrule
1 & $\begin{array}{c}X_{5/3} \bar u_i \\ \bar Y_{-4/3} d_i \end{array}$ & $u_i \bar u_i + W^+$ & $d_i \bar d_i + W^+$ \\
\midrule
4/3 & $X_{5/3} d_i$ & $u_i d_i + W^+$ & \\
\toprule
\end{tabular}
\caption{Production and decay channels for $X_{5/3}$ and $Y_{-4/3}$. Only configurations with $Q\geq0$ have been shown for simplicity. Throughout the table, $u_i=u,c,t(,t^\prime)$ and $d_i=d,s,b(,b^\prime)$.}
\label{XYproduction}
\end{table}

VLQs with exotic charges can only interact with other states through charged currents. They can be produced singly or in pair, but the allowed diagrams and decay channels are limited, with respect to the $\{t^\prime,b^\prime\}$ case, due to the smaller number of interactions. VLQs with exotic charges always belong to multiplets which contain also one SM VL partner (non-SM doublets) or two SM VL partners (triplets), therefore they can be produced together with a $t^\prime$ or $b^\prime$. All the possible combinations are listed in Tab.~\ref{XYproduction}. 

The allowed decays for the $X_{5/3},Y_{-4/3}$ VLQs are:
\begin{eqnarray}
 X_{5/3}  \to u_i W^+ &\mbox{ with }& u_i=u,c,t,t^\prime \\
 Y_{-4/3} \to d_i W^- &\mbox{ with }& d_i=d,s,b,b^\prime 
\end{eqnarray}
Again, not all possibilities may be allowed for a given scenario, especially decays into $t^\prime$ and $b^\prime$, which depend on the mass gaps inside the multiplets. The variation of the branching ratios for a $X_{5/3}$ belonging to the non-SM doublet $(X_{5/3},t^\prime)$ are shown in Fig.~\ref{BRstprimeX}. It is still possible to notice that assuming mixing only with third family can be a quite strong assumption, which may lead to misinterpreting experimental results and overestimating mass bounds.

\subsection{Results from specific phenomenological analyses}

LHC signatures of VLQs have been analysed in many phenomenological studies, both in specific scenarios and in model-independent ways. In most analyses it is commonly assumed that VLQs mix only with third generation quarks, while some analyses consider the general mixing case. 
If VLQs mix mostly with third generation quarks a reinterpretation of many experimental searches for top or bottom partners, even under explicit chiral assumptions, is possible, since the favourite channels involve the presence of top quark decay products, which can be obtained also in VLQs decays. 
The scenario in which the VLQs do not mix with at all with the third generation, but only with lighter generations, has also been considered, and its main advantage is the presence of energetic jets in the final state, due to decays of the type $Q\to Vj$. This scenario has been searched experimentally, and therefore predictions can be directly compared with data.
Finally, for scenarios with general mixing, a reinterpretation of experimental bounds is not straightforward since acceptances for different channels may differ significantly. 

In the following, a brief overview of phenomenological analyses is presented. The large amount of studies in literature makes it impossible to analyse all of them in a single review, therefore a selection must be made, considering only:
\begin{itemize}
\item studies which have been published on peer reviewed journal at the time of submission of the present review;
\item analyses of minimal scenarios, i.e. signatures coming from single or pair production of VLQs, which then directly decay into SM states. This choice has been made to keep the overview as model independent as possible, and to be as close as possible to the minimal framework described in Sect.~\ref{Sect:model}; furthermore, model-independent predictions can be easily compared with current experimental searches.
\end{itemize}
The analyses are presented in chronological order; descriptions of the main assumptions on VLQ properties and of the proposed discovery channels are provided, but more details can be found in the original publications.\\

In \cite{Contino:2008hi} the pair production of $X_{5/3}$ and of $b^\prime$ is considered. The heavy states are then required to decay 100\% into $W^\pm t$, depending on the case. The analysis then focuses on signatures with same sign dileptons in the final state, i.e.:
\begin{equation}
 p p \to X_{5/3}\bar X_{5/3} , b^\prime \bar b^\prime \to l^\pm \nu l^\pm \nu b \bar b q \bar q q \bar q
\end{equation}
The detailed analysis shows how the same-sign dilepton channel can be extremely promising for the discovery of heavy states which mainly mix to third family quarks. The same-sign dilepton channel has been considered also in \cite{Mrazek:2009yu,Dissertori:2010ug} where method for the reconstruction of the $X_{5/3}$ mass are proposed.

In \cite{AguilarSaavedra:2009es} a thorough systematic analysis of signatures of all possible VLQs states, mixing mainly with third generation, is undertaken. All singlet and doublet scenarios are analysed and benchmark points for mixing parameters are considered. Final states with different number of leptons are combined to obtain discovery potentials for each state at LHC with a center of mass energy of 14 TeV . 

In \cite{Atre:2011ae} the single production of VL quarks through the process $q q^\prime \stackrel{V^*}{\to} q_1Q$ with $V=Z,W$ is considered for LHC at both 7 and 14 TeV. The analysis is performed factorizing out the model-dependence in the coupling parameters and considering $Q=X_{5/3},t^\prime,b^\prime,Y_{-4/3}$. The VLQs are assumed to decay through CC and NC to $Vj$, but decays to $hj$ are not considered, and to reduce the background, leptonic decays of the gauge bosons are considered. The considered processes are therefore:
\begin{eqnarray}
 p p \to (Q \to Vj)j \quad\mbox{with } V\to \{l^\pm l^\mp, l^\pm \MET, \MET \}
\end{eqnarray}
Different scenarios, with constraints on coupling parameters are considered. The analysis shows that despite in the first 7 TeV run with $1~\mathrm{fb}^{-1}$ luminosity the reach of the LHC is similar to the Tevatron reach with a $10~\mathrm{fb}^{-1}$ luminosity, the reach greatly improves in the long run configuration with $\sqrt{s}=14$ TeV and $L=100~\mathrm{fb}^{-1}$, with the possibility to probe masses as high as $\sim 3.7$ TeV.

In \cite{Gopalakrishna:2011ef} the pair and single production of $b^\prime$ at the LHC with a center of mass energy of 14 TeV is studied, focusing on $Zb$ and $bH$ decays with leptonic and semileptonic final states. The needed luminosity for discovery of a $b^\prime$ is plotted against the $b^\prime$ mass up to 1250 GeV.

In \cite{Cacciapaglia:2011fx} signatures at LHC with $\sqrt{s}=7$ TeV for single production of a $t^\prime$ belonging to a non-SM VLQ doublet $\{X_{5/3},t^\prime\}$ are studied. The processes are studied for $m_{t^\prime}=350-500$ GeV, a range around the value for which single production becomes dominant over pair production. FCNCs decays to $Zt$ and $th$ are considered, but since the $t^\prime$ is allowed to mix with all SM families, relations between BRs for different masses have been considered. The analysis focuses on the following signatures:
\begin{eqnarray}
 p p \to (t^\prime \to Zt) j &\quad&\mbox{with } Z \to l^+l^-, \nu \bar \nu, jj \\
 p p \to (t^\prime \to th) j &\quad&\mbox{with } h \to b\bar b   
\end{eqnarray}
Also, a signature under the hypothesis of no mixing with third family is considered, namely:
\begin{equation}
 p p \to (t^\prime \to jh) j \quad\mbox{with } h \to b \bar b
\end{equation}
The analysis show that such signatures are difficult to observe at LHC with a center of mass energy of 7 TeV, since SM backgrounds have very similar kinematical features.

In \cite{Azatov:2012rj} signatures at LHC with $\sqrt{s}=14$ TeV for pair production of a VL singlet labelled $t_2$, in a model which contains also a SM-like VL doublet, are studied. One of the $t_2$ is required to decay to $th$, while the other is free to decay to $Zt$, $th$ or $Wb$. Depending on the considered Higgs mass (125 or $>$200 GeV), Higgs decays $h\to\gamma\gamma$ or $h\to ZZ\to4l$ are considered. The processes under consideration are therefore:
\begin{eqnarray}
 p p \to t_2 \bar t_2 \to \left\{\begin{array}{l} thth \\ thZt \\ thWb \end{array}\right. \quad\mbox{with } h \to \gamma\gamma \mbox{ or } h\to ZZ\to4l 
\end{eqnarray}
The analysis shows that these channel can be observed at $14$ TeV for a luminosity between $\sim 20~\mathrm{fb}^{-1}$ and $\sim 200~\mathrm{fb}^{-1}$ for $120~\mathrm{GeV}\lesssim m_h \lesssim 550~\mathrm{GeV}$ and the definition of new variables is proposed, to allow the estimation of the heavy top mass. 

In \cite{Harigaya:2012ir} processes of production and decay of a singlet top partner are considered for LHC with a center of mass energy of 8 TeV and integrated luminosity of $15~\mathrm{fb}^{-1}$. The study focuses on the $tH$ decay channel, which is enhanced by the presence of a dimension-5 operator in the effective lagrangian. Considering final states with multiple b-jets, bounds on the top partner mass and mixing angles parameter space are provided.

\section{Contribution of vector-like quarks to other processes at LHC}
\label{Sect:VLQcontributions}

\subsection{Higgs production and decay}

VLQs can contribute to processes of Higgs production and decay: $t^\prime$ and $b^\prime$ quarks mix with SM states through Yukawa couplings, and therefore they can circulate in the loops of gluon fusion processes and $h\to\gamma\gamma$ decays, as depicted in Fig.~\ref{VLQforotherproc}. In minimal pictures, adding only one VLQ representation, exotic VLQs do not play any role at LO, but can provide corrections to these processes at higher orders. Many studies have dealt with the role of VLQs to Higgs production processes \cite{Cacciapaglia:2011fx,Azatov:2011qy}, finding that adding just one VL multiplet the corrections to gluon fusion and $\gamma\gamma$ decays are always negligibly small. The same result applies also at NNLO \cite{Dawson:2012di}. 

Other studies have considered Higgs signatures given by VLQ decays, which can be considered together with standard Higgs production and decay channels.
In \cite{Kribs:2010ii} processes of Higgs production driven by the decay of a $t^\prime$ singlet are considered. The $t^\prime$ is allowed to mix only with the top quark and the analysis focuses on pair production at 14 TeV, where one of the two top partners decays to $th$ and the other into $Zt$ or $Wb$. The search is then conducted looking for boosted Higgs bosons which then decay into $b \bar b$. Though lacking a realistic detector simulation, the analysis describes a scenario in which the Higgs signal can be enhanced by the presence of $t^\prime$ in the mass range 400-800 GeV. In this framework, the Higgs can thus be observed with a 5$\sigma$ significance at a luminosity of 10 fb$^{-1}$.

In \cite{Asakawa:2010xj} corrections from new physics to the trilinear Higgs coupling and to Higgs pair production are studied. In the case of VLQs, an example with a singlet top partner is considered: the analysis shows that the contribution of VLQs is small, of the order of few percent, and this is due to decoupling properties of the heavy state.

\subsection{Same-sign top production}

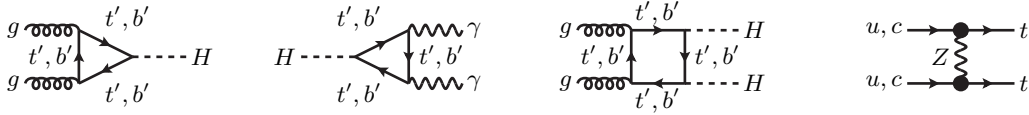
\begin{figure}[htb]
\begin{center}
 \begin{picture}(100,40)(0,-10)
  \SetWidth{1}
  \Gluon(10,20)(30,20){2}{4}
  \Text(8,20)[rc]{\small{$g$}}
  \Gluon(10,0)(30,0){2}{4}
  \Text(8,0)[rc]{\small{$g$}}
  \Line[arrow,arrowpos=0.5,arrowlength=2,arrowwidth=1,arrowinset=0.2](30,0)(30,20)
  \Text(28,10)[rc]{\small{$t^\prime,b^\prime$}}
  \Line[arrow,arrowpos=0.5,arrowlength=2,arrowwidth=1,arrowinset=0.2](30,20)(50,10)
  \Text(40,20)[lb]{\small{$t^\prime,b^\prime$}}
  \Line[arrow,arrowpos=0.5,arrowlength=2,arrowwidth=1,arrowinset=0.2](50,10)(30,0)
  \Text(40,0)[lt]{\small{$t^\prime,b^\prime$}}
  \Line[dash,dashsize=2.5](50,10)(70,10)
  \Text(72,10)[lc]{\small{$H$}}
 \end{picture}
 \begin{picture}(100,40)(0,-10)
  \SetWidth{1}
  \Line[dash,dashsize=2.5](10,10)(30,10)
  \Text(8,10)[rc]{\small{$H$}}
  \Line[arrow,arrowpos=0.5,arrowlength=2,arrowwidth=1,arrowinset=0.2](30,10)(50,20)
  \Text(40,20)[rb]{\small{$t^\prime,b^\prime$}}
  \Line[arrow,arrowpos=0.5,arrowlength=2,arrowwidth=1,arrowinset=0.2](50,20)(50,0)
  \Text(54,10)[lc]{\small{$t^\prime,b^\prime$}}
  \Line[arrow,arrowpos=0.5,arrowlength=2,arrowwidth=1,arrowinset=0.2](50,0)(30,10)
  \Text(40,0)[rt]{\small{$t^\prime,b^\prime$}}
  \Photon(50,20)(70,20){2}{4}
  \Text(72,20)[lc]{\small{$\gamma$}}
  \Photon(50,0)(70,0){2}{4}
  \Text(72,0)[lc]{\small{$\gamma$}}
 \end{picture}
 \begin{picture}(100,40)(0,-10)
  \SetWidth{1}
  \Gluon(10,20)(30,20){2}{4}
  \Text(8,20)[rc]{\small{$g$}}
  \Gluon(10,0)(30,0){2}{4}
  \Text(8,0)[rc]{\small{$g$}}
  \Line[arrow,arrowpos=0.5,arrowlength=2,arrowwidth=1,arrowinset=0.2](30,0)(30,20)
  \Text(28,10)[rc]{\small{$t^\prime,b^\prime$}}
  \Line[arrow,arrowpos=0.5,arrowlength=2,arrowwidth=1,arrowinset=0.2](30,20)(50,20)
  \Text(40,22)[cb]{\small{$t^\prime,b^\prime$}}
  \Line[arrow,arrowpos=0.5,arrowlength=2,arrowwidth=1,arrowinset=0.2](50,20)(50,0)
  \Text(54,10)[lc]{\small{$t^\prime,b^\prime$}}
  \Line[arrow,arrowpos=0.5,arrowlength=2,arrowwidth=1,arrowinset=0.2](50,0)(30,0)
  \Text(40,-2)[ct]{\small{$t^\prime,b^\prime$}}
  \Line[dash,dashsize=2.5](50,20)(70,20)
  \Text(72,20)[lc]{\small{$H$}}
  \Line[dash,dashsize=2.5](50,0)(70,0)
  \Text(72,0)[lc]{\small{$H$}}
 \end{picture}
 \begin{picture}(100,40)(-20,-10)
  \SetWidth{1}
  \Line[arrow,arrowpos=0.5,arrowlength=2,arrowwidth=1,arrowinset=0.2](10,20)(30,20)
  \Text(8,20)[rc]{\small{$u,c$}}
  \Line[arrow,arrowpos=0.5,arrowlength=2,arrowwidth=1,arrowinset=0.2](30,20)(50,20)
  \Text(52,20)[lc]{\small{$t$}}
  \Line[arrow,arrowpos=0.5,arrowlength=2,arrowwidth=1,arrowinset=0.2](10,0)(30,0)
  \Text(8,0)[rc]{\small{$u,c$}}
  \Line[arrow,arrowpos=0.5,arrowlength=2,arrowwidth=1,arrowinset=0.2](30,0)(50,0)
  \Text(52,0)[lc]{\small{$t$}}
  \Photon(30,20)(30,0){2}{4}
  \Text(26,10)[rc]{\small{$Z$}}
  \Vertex(30,20){3}
  \Vertex(30,0){3}
 \end{picture}
\end{center}
\caption{Feynman diagrams for processes with VLQs contributions. From left to right: Higgs production through gluon fusion, Higgs to two photon decay, Higgs pair production, FCNC same-sign top production.}
\label{VLQforotherproc}
\end{figure}

The process of production of same-sign tops at the LHC can be driven by the presence of VLQs due to the possibility of FCNCs. The Feynman diagram is shown in Fig.~\ref{VLQforotherproc}. The analysis in \cite{Lee:2004me} shows that, in a Littlest Higgs scneario with one top-partner VL quark, the cross section is however very small, around 0.01 fb at LHC with a center of mass energy of 14 TeV, below the experimental sensitivity.

\section{Conclusion}

The aim of this review has been to provide a broad, though necessarily incomplete, overview about the searches and perspectives of heavy vector-like quarks at the LHC. Vector-like quarks are predicted by many models of new physics. Recent observations strongly point towards the existence of the Higgs boson, thus completing the SM picture: among the next steps of the LHC, there will be therefore the search for new BSM states. A minimal extension of the SM with the presence of vector-like quarks has a huge and interesting range of possible signatures, some of which have already been tested both at Tevatron and at the LHC. Current bounds on the mass of vector-like quarks are around 400-600 GeV, depending on assumptions on their mixing and decay channels. If vector-like quarks mix with all SM families many searches must be reinterpreted while dedicated, optimized, searches may be in order. A complete list of possible final states for production of any possible vector-like quark in the minimal picture has been 
provided, together with a short description of the main phenomenological analyses present in literature. The discovery of a new fermionic state would certainly be a major and exciting event at the LHC, thus a detailed understanding of its properties, if it turn out to be vector-like, will be extremely useful for future analyses.

\section*{Acknowledgements}
The research of Y.O. is supported in part by the
Grant-in-Aid for Science Research, Japan Society for the Promotion of
Science (JSPS), No. 20244037 and No. 22244031.

\end{document}